\documentclass[aps,11pt,showkeys,nofootinbib]{revtex4-1}

\usepackage{amsmath,amssymb,amsfonts,amsthm}
\usepackage{amsbsy} 
\usepackage{epsfig}
\usepackage{latexsym}

\usepackage[dvipsnames]{xcolor}
\usepackage{soul} 
\usepackage{subcaption}
\usepackage{braket}
\usepackage{tensor}
\usepackage{ulem}

\input amssym.def
\input amssym.tex

\usepackage{hyperref}

\newcommand{\mk}[0]{\mathcal{K}}
\newcommand{\mr}[0]{\mathcal{R}}

\newcommand{\ma}[0]{\mathcal{A}}
\newcommand{\mb}[0]{\mathcal{B}}

\newcommand{\mmp}{\mathcal{P}}

\newcommand{\tPhi}[0]{\Tilde{\Phi}}

\newcommand{\sgn}[1]{\text{sgn}(#1)}

\newcommand{\mkz}{\mathcal{K}^{(0)}}
\newcommand{\mkt}{\mathcal{K}^{(2)}}
\newcommand{\lagr}{\mathcal{L}}

\newcommand{\chiz}{(\chi^0)}

\newcommand{\expv}[1]{\langle #1 \rangle}
\newcommand{\ev}[1]{\langle #1 \rangle}

\newcommand{\piInt}[1]{\int \frac{{\rm d}^3 #1}{(2 \pi)^3}}

\newcommand{\pd}{\partial}

\newcommand{\ii}{\text{i}}

\newcommand{\nk}{\mathfrak{n}_k(\chi^0)}
\newcommand{\ck}{\mathfrak{c}_k}
\newcommand{\mko}{\mathfrak{f}_k(\chi^0)}
\newcommand{\dk}{\mathfrak{d}_k}

\newcommand{\no}[1]{:#1\!:}

\newcommand{\refsec}{sec.\,}

\newcommand{\refapp}{app.\,}

\def\be{\begin{equation}}
\def\ee{\end{equation}}
\def\dd{{\rm d}}
\def\bes{\begin{eqnarray}}
\def\ees{\end{eqnarray}}
\def\T{{\mathcal{T}}}

\begin{document}

\title{
Cosmological scalar perturbations for a metric reconstructed from group field
theory
}
\author{Steffen Gielen}
\email{s.c.gielen@sheffield.ac.uk}
\affiliation{School of Mathematical and Physical Sciences, University of Sheffield, Hicks Building, Hounsfield Road, Sheffield S3 7RH, United Kingdom}
\author{Lisa Mickel}
\email{lisa.mickel@iap.fr}
\affiliation{Institut d'Astrophysique de Paris, 98 bis Boulevard Arago,  75014 Paris, France}
\affiliation{School of Mathematical and Physical Sciences, University of Sheffield, Hicks Building, Hounsfield Road, Sheffield S3 7RH, United Kingdom}
\date{\today}

\begin{abstract}

While homogeneous cosmologies have long been studied in the group field theory (GFT) approach to quantum gravity, including a quantum description of cosmological perturbations is highly non-trivial. Here we apply a recent proposal for reconstructing an effective spacetime metric in GFT to the case of a metric with small inhomogeneities over a homogeneous background. We detail the procedure and give general expressions for cosmological scalar perturbations defined in terms of the GFT energy-momentum tensor. These include all the scalar components of standard perturbation theory and hence can be used to define gauge-invariant quantities. 
This is a major advantage of the effective metric approach compared to previous GFT studies limited to volume perturbations.
We compute these perturbations explicitly for a particular Fock coherent state. While it was previously shown that such a state can be interpreted as an approximately flat homogeneous cosmology at late times, here we find that, in a very simple example,
inhomogeneities do not follow the dynamics of general relativity in the semiclassical regime. 

More specifically, restricting ourselves to a specific coherent state in a simple (free) GFT, we study two types of perturbative GFT modes, squeezed and oscillating modes. For squeezed modes we find perturbation equations with Euclidean signature and a late-time limit that differs from general relativistic perturbation equations. Oscillating modes satisfy different dynamical equations that also differ from those of general relativity, but show a Lorentzian signature. Considering that our results were obtained within a number of simplifying assumptions and arguably the simplest possible example, we discuss how going beyond these assumptions could lead to a more desirable phenomenology. Overall, our analysis should be understood as a first step in understanding cosmological perturbations within the effective GFT metric. 

\end{abstract}

\maketitle

\section{Introduction}

General relativity gives an excellent classical description of the gravitational force; however, the quest to find a quantum theory of gravity is still ongoing. 
The quantum behaviour of matter is well described by quantum field theory and as general relativity relates matter to the geometry of spacetime, it is generally believed that a fully consistent theory also requires a quantum description of geometry (though see \cite{Oppenheim:2018igd} for an alternative viewpoint). Quantum gravity is also expected to cure the singularities through which classical general relativity predicts its own incompleteness.

Finding a satisfactory quantum description of gravity is no easy feat. The perhaps natural approach of applying quantum field theory techniques to a metric perturbation around a Minkowski background leads to a non-renormalisable theory \cite{Goroff:1985th}.
Multiple approaches to finding a quantum formulation of gravity have been established \cite{Oritibook,whitepaper}; different approaches have vastly different starting points and it is not necessarily clear how and if they connect to each other.
It is often difficult to carry out explicit calculations within quantum gravity to assess if and in which way a given theory relates to general relativity in a suitable classical limit. A frequent strategy to circumvent this issue and obtain first insights into the physical viability and implications of a specific approach is its application to the cosmological setting, where the high degree of symmetry significantly reduces the relevant number of degrees of freedom. 
Within general relativity, homogeneity and isotropy of our cosmos are captured by the Friedmann--Lema\^{i}tre--Robertson--Walker (FLRW) metric; in standard cosmology the universe is modelled as a (flat) FLRW spacetime with small inhomogeneous perturbations. 
In addition to obtaining a description for the background metric,  it is desirable to include a description of cosmological perturbations within the quantum framework to make further contact with general relativity and possibly even cosmological observations. 

Here, we work within group field theory (GFT)  \cite{Freidel:2005qe, Oriti:2011jm}, a background-independent approach to quantum gravity related to loop quantum gravity (LQG) \cite{pullin, Ashtekar_2021, bodendorfer} and spin foam models \cite{Engle_2023}.
GFTs first appeared in the form of a 3-dimensional quantum gravity model \cite{BOULATOV_1992}; they have been studied in the context of models related to LQG and spin foam models \cite{Oriti:2011jm, Oriti:2013aqa, DePietri:1999bx, Reisenberger:2000zc,Ben_Geloun_2010} and developed into their own research field.
A GFT is a field theory defined on an abstract group manifold; hence 
GFT does not presuppose a spacetime manifold, but spacetime is dynamically emergent from a large number of GFT quanta, which should be understood as the building blocks of space. 
This picture is sometimes illustrated with an analogy to fluid dynamics where a large collection of water molecules (GFT quanta) leads to the emergence of a fluid (spacetime), which is characterised by different attributes than the single molecules and described by different dynamical laws \cite{Oriti:2007qd}. One is then led to the idea of a macroscopic universe emerging from a ``condensate'' of GFT quanta \cite{sindoniReview, Oriti_2017, universe5060147}, described by a coherent many-body quantum state similar to those appearing in condensed matter physics. Using this main idea, the
application of GFT to effective FLRW geometries (modelled by a simple coherent quantum state) shows a resolution of the Big Bang singularity, which is replaced by a bounce that interpolates between a contracting and an expanding phase \cite{GFTscalarcosmo, Oriti:2016ueo}. 
Extensions of this scenario can introduce additional interesting phenomenological features in the cosmological evolution \cite{cyclicUniverse, Oriti:2021rvm, Calcinari:2022iss}. 
Phenomenologically interesting homogeneous cosmologies can be obtained from a broad range of underlying GFT models.  
In order to further establish GFT as an approach to quantum gravity, it is however imperative to study its implications beyond this rather restricted setting of homogeneous cosmology. A first and natural extension is then to consider inhomogeneous cosmological perturbations.

A basic idea proposed in previous work on GFT cosmology is that massless scalar fields are used to define ``relational'' coordinates. To describe homogeneous cosmology, a single matter field is sufficient, whereas for a treatment of inhomogeneities one usually includes four matter fields that can form a relational coordinate system. 
The concept of relational coordinates has been widely investigated within general relativity as a means to define local gauge-invariant observables (see, e.g., \cite{Rovelli:1990ph,Giddings:2005id, Dittrich:2005kc, Tambornino:2011vg}). 
The idea is to construct an observable 
 by considering the value of a quantity of interest at the spacetime point defined by the value of another physical quantity, which can be a diffeomorphism-invariant definition unlike the standard coordinate-dependent tensorial quantities.
 In GFT, models with a single clock field have been applied to homogeneous cosmology since the proposal of \cite{GFTscalarcosmo,Oriti:2016ueo}; models with four (or more) fields have been introduced more recently \cite{Gielen:2017eco,Gerhardt:2018byq,Gielen_2018,Gielen_CMB,Marchetti:2021gcv,Jercher:2023nxa, Jercher_2024_lett}. 
We will use the same construction of four massless scalar fields as relational coordinates. While this is not a realistic model of cosmology, and more work would be needed to connect to scenarios such as inflation, the initial goal of this line of research is to establish whether predictions of GFT in this setting are compatible with those of general relativity with a similar matter content.

Information about the emergent spacetime in GFT can be extracted from expectation values of relevant operators in suitably semiclassical states, where semiclassicality is a necessary criterion for emergence of a classical spacetime in the multiparticle limit \cite{sindoniReview, deparamcosmo, Calcinari:2023sax}. 
Previous GFT literature predominantly makes use of the volume operator, based on the assumption that volume eigenvalues of GFT quanta are given by the eigenvalues of the LQG volume operator \cite{Rovelli:1994ge,Brunnemann_2006} for comparable spin-network vertices.
In this approach, the main observable used to compare with the classical cosmology is the total volume as a function of a matter clock. In this paper, we deviate from this conventional approach and make use of the proposal to reconstruct an effective metric from GFT operators detailed in \cite{gftmetric}.
This proposal relies on the identification of Noether currents in the classical theory with expectation values of corresponding GFT operators. As the spacetime metric contains more information than just the volume of a spacetime region, this new approach potentially gives access to a wider class of observables (including vector and tensor modes which do not appear in the volume). 
The access to additional properties of spacetime is the main point of attractiveness of developing the effective metric approach to extract semiclassical quantities. 

In usual spacetime physics, the action of four massless scalar fields that span a relational coordinate system exhibits a shift symmetry; the same symmetry is imposed when these fields are introduced in GFT, leading to the above-mentioned Noether currents. 
Specifically, this symmetry allows the definition of a conserved GFT energy-momentum tensor in analogy to the energy-momentum tensor of standard field theories. 
The expectation value of the GFT energy-momentum tensor is then identified with the classical Noether currents arising from the shift symmetry. 
In the relational coordinate system the classical Noether currents are related to the components of the metric, and the conservation law for these currents is the Klein--Gordon equation for the matter fields.
An \emph{effective} metric can then be reconstructed from the expectation values of the operators corresponding to the GFT energy-momentum tensor \cite{gftmetric}, where ``effective'' refers to the fact that the metric is obtained from operator expectation values over semiclassical states and there is no corresponding metric \emph{operator} at the quantum level.

In \cite{gftmetric} we explored the application of the effective metric proposal to a flat FLRW cosmology, studying the homogeneous mode of the GFT energy-momentum tensor in a Fock coherent state (defined in a way that is similar to previous GFT literature, e.g., \cite{deparamcosmo,Calcinari:2023sax}).
We showed that the resulting metric leads to a bounce and can be consistently interpreted as a flat FLRW metric in the semiclassical regime away from the bounce. 
However, the effective Friedmann equation one recovers at late times  disagrees with general relativity coupled to four massless scalar fields, and rather corresponds to the equation expected for only a single massless scalar field. 

Here, we extend this analysis to inhomogeneous modes interpreted as cosmological scalar perturbations. Our work connects to previous studies of cosmological perturbations in GFT, which also rely on a relational coordinate system spanned by four massless scalar fields \cite{Gielen:2017eco,Gielen_2018, Gielen_CMB, Marchetti:2021gcv,Jercher:2023nxa, Jercher_2024_lett}. 
More specifically, the previous works \cite{Marchetti:2021gcv,Jercher:2023nxa, Jercher_2024_lett} study perturbations in a scenario in which the effective GFT Friedmann equation also agrees with that of general relativity with a single scalar field.
Those works include a fifth matter field that is then assumed to be the dominant matter content of the universe, which could justify the GFT Friedmann equation. In this paper, we work in a setting with four fields all appearing on the same footing, and do not explicitly address the discrepancy at background level. 
This issue is also mentioned in \cite{gftmetric}.

Previous investigations consider the volume operator and its perturbations, which restricts the perturbative quantities that can be studied; this limitation is absent when using a GFT effective metric. In our approach, all perturbative quantities (including gauge-invariant ones) can
be reconstructed from the effective metric; in this paper we limit our study to scalar perturbations but analogous constructions for vector and tensor modes should be possible.
We give general relations between GFT operator expectation values and scalar metric perturbation variables that arise directly from the effective GFT metric proposal and hold for any choice of semiclassical GFT state. 
We then find concrete expressions for scalar perturbation variables for the state used already in \cite{gftmetric}.
The perturbative dynamics of these variables agree neither with those of general relativity with four scalar fields nor with those of general relativity with one scalar field, as might be suggested by the background dynamics. 
In particular, the dominant squeezed GFT modes exhibit exponentially growing behaviour, as one would expect from Euclidean rather than Lorentzian signature. This is a more fundamental type of disagreement with the general relativistic dynamics than a choice of matter content as in the case of the background dynamics. 
We hope that the analysis presented can serve as a blueprint for related studies that might consider different state choices or amendments to the underlying GFT model. For instance, one could compare with more sophisticated constructions in the GFT literature that can lead to phenomenologically more acceptable results for perturbations, such as  \cite{Jercher:2023nxa, Jercher_2024_lett}.

Our results open up an avenue to studying gauge-invariant quantities and more general perturbation variables, surpassing the limitations of previous work restricted to the perturbed volume element only. Extending the setup to give phenomenologically more realistic results then has the potential to connect GFT to observables relevant to cosmological observations. 
Within the various assumptions we have made, the GFT effective metric calculations do not reproduce perturbative dynamics compatible with general relativity; therefore, either our various simplifying assumptions (such as limiting to the free theory and the particular choice of state) are not all justified or the effective metric construction or the particular class of GFT models is ruled out. 
The hypothetical opposite result, a calculation showing agreement of GFT with general relativity after neglecting GFT interactions, for the simplest coherent state choice, and for an arbitrary choice of (compact) gauge group, should indeed be seen as highly implausible.
Our work should be understood as a first step in the challenging task of gaining further insights into the phenomenology of GFT.

Let us emphasise some conceptual differences between the approach taken here and that of standard cosmological perturbation theory \cite{cosmopert} or loop quantum cosmology (LQC) \cite{Agullo_2023}. 
In these standard approaches, the background and perturbations are treated as separate entities. For instance, in LQC perturbations can be included by quantising the background and perturbations separately \cite{Agullo_2013_dressedMetric}, such that the perturbations evolve on an effectively classical background with LQC corrections, or by working in an effective framework and ensuring that the algebra of the modified constraints is anomaly-free \cite{LQCanomaly}.
In contrast, our effective metric approach in GFT treats the background and perturbations on the same footing as they simply correspond to different wavenumbers. As a consequence, the perturbations are fully treated within GFT and not quantised with different methods or on a pre-existing background. They are also both determined by a single quantum state, with less freedom to set arbitrary initial conditions.

This paper is structured as follows.
We first outline the main ideas of the GFT framework and the specific formulation we utilise in this paper in \refsec \ref{sec:GFT}.
In \refsec \ref{sec:GFTmetric_construction} we review the main premise of the effective GFT metric which is built on the idea of a conserved GFT energy-momentum tensor as first introduced in \cite{gftmetric}.
We establish the relation between expectation values of the GFT energy-momentum tensor components and the perturbed flat FLRW metric in \refsec \ref{sec:GFTuniverse_effMetric}.
Sec.\,\ref{sec:classicalAnalysis} is dedicated to the analysis of a classical perturbed FLRW spacetime in a relational coordinate system spanned by four massless scalar fields, which, while straightforward in principle, is not commonly discussed in the literature. 
In \refsec \ref{sec:GFTuniverse_background} we discuss our choice of state that reflects the required symmetries of the cosmological setting and revise the effective background metric arising from the homogeneous background mode, where it was shown in \cite{gftmetric} that the recovered metric is flat and gives a bouncing universe. 
In \refsec \ref{sec:GFTuniverse_perturbations} we extend the past analysis to cosmological perturbations.
After general considerations, we include detailed calculations for squeezed and oscillating modes. 
We conclude in \refsec \ref{sec:conclusion}.
Possible extensions to our setup are discussed in the appendix. 
New results are contained in \refsec\ref{sec:GFTuniverse_effMetric}, \ref{sec:classicalAnalysis_perturbations}, and \ref{sec:GFTuniverse_perturbations}; other sections briefly review the results of \cite{gftmetric} in order to make this paper as self-contained as possible.

\section{Elementary aspects of Group Field Theory}
\label{sec:GFT}

GFT is a background-independent approach to quantum gravity in which spacetime emerges from the excitations of an abstract quantum field defined on a group manifold. The group manifold is not thought of as spacetime but as a configuration space for discrete gravity and matter. We direct the interested reader to \cite{Freidel:2005qe,Oriti:2011jm,Baratin:2011aa} for reviews as we can only sketch the main ideas here.

A GFT model for quantum gravity in vacuum can be defined in terms of a group field $\varphi(g_i)$ (here chosen to be real-valued) and an action $S[\varphi]$. The arguments of $\varphi$ are $n$ group elements $g_i$ ($i = 1,   \ldots, n$) valued in a suitable gauge group, with $n$ usually representing the expected spacetime dimension.  
Schematically (see, e.g., \cite{Oriti:2011jm})
we expand the partition function perturbatively as
\be
Z = \int \mathcal{D}\varphi\;e^{-S[\varphi]}=\sum_\Gamma \frac{\lambda^{v_\Gamma}}{{\rm sym}[\Gamma]}A[\Gamma]\,,
\label{eq:GFTZ}
\ee
where $\Gamma$ are Feynman graphs, ${\rm sym}[\Gamma]$ is a symmetry factor and $A[\Gamma]$ a Feynman amplitude. Here we are assuming a single interaction term in $S[\varphi]$ with coupling $\lambda$, and $v_\Gamma$ is the number of interaction vertices in $\Gamma$. For a suitably chosen action, $A[\Gamma]$ represents a discrete quantum gravity path integral or spin foam amplitude associated to the graph $\Gamma$, which is interpreted as a combinatorially defined discrete spacetime. 
The boundary states of the graph correspond to (triangulations of) spatial hypersurfaces.
In the example of the Boulatov model \cite{BOULATOV_1992}, the Feynman graphs represent oriented three-dimensional simplicial complexes and $A[\Gamma]$ would be the amplitude defining the Ponzano--Regge model of three-dimensional quantum gravity \cite{ponzanoregge}. In this case, we have a field $\varphi(g_1,\, g_2, \, g_3)$ defined on ${\rm SU} (2)^3$, where ${\rm SU}(2)$ corresponds to the gauge group for gravity. In general, $Z$ would define a sum over all possible discrete spacetime histories weighted by path integral-like amplitudes. Just as it would be the case for a more conventional quantum field theory, the type of kinematical data associated to each $\Gamma$ depends on the kinematical data chosen for the group field (in particular, on the choice of gauge group), while the precise choice of interaction term(s) determines what types of graphs $\Gamma$ appear in the sum, and what amplitudes are associated to each $\Gamma$.

In the present context, we are interested in models for four-dimensional quantum gravity coupled to four scalar matter fields. We will later use these matter fields to define a relational coordinate system. 
In addition to four group arguments $g_i$ ($i = 1, \, 2, \, 3, \, 4$), which we will here choose to be elements of ${\rm SU}(2)$, we therefore couple four scalar fields $\chi^A$ ($A = 0, \, 1, \, 2, \, 3$) to the group field $\varphi(g_i, \chi^A)$, which becomes a function $\varphi: \rm{SU}(2)^4 \times \mathbb{R}^4 \to \mathbb{R}$.
Such a field can be expanded in modes associated to ${\rm SU}(2)$ representations,
\be
\varphi(g_i,\chi^A) = \sum_J \varphi_J(\chi^A)\,D_J(g_i)\, ,
\label{eq:modeDecomp}
\ee
where $D_J(g_i)$ represent suitable combinations of Wigner $D$-matrices and $J = (\Vec{j},\, \Vec{m},\, \iota)$ is a multi-index representing ${\rm SU}(2)$ irreducible representations $\Vec{j} = (j_1, \, j_2, \, j_3, \, j_4)$, the corresponding magnetic indices $\Vec{m}$ with $m_i \in \{ -j_i,\, -j_i + 1 \ldots, j_i -1, \, j_i \}$, and intertwiners $\iota$, which label the basis of the subspace invariant under $\text{SU}(2)$ transformations. Since the multi-index $J$ will remain abstract in the following, an equivalent construction would be possible for any model with compact gauge group allowing for a similar mode expansion (for instance, if we chose ${\rm U}(1)$ the different $J$ would just be discrete Fourier modes on the circle). In this sense, our formalism is very general. One might want to extend the construction to non-compact groups such as $\text{SL}(2,\mathbb{C})$, which is used in some Lorentzian four-dimensional models \cite{Jercher:2021bie,Jercher:2022mky} and could be seen as the natural gauge group in that context, even though there are also four-dimensional Lorentzian models based on $\text{SU}(2)$ (see, e.g., \cite{Marchetti:2021gcv} for a comparison). In the non-compact case, (\ref{eq:modeDecomp}) needs to be replaced by a more complicated expression involving integrals over continuous representation labels. We will later return to the question of whether the spacetime signature is indeed encoded in a choice of GFT gauge group.

In general, the GFT action contains a quadratic part and higher order interactions nonlocal in the group arguments,  defined in a specific way to give Feynman graphs the desired combinatorial structure to be interpreted as spacetime histories, as we have discussed. The quadratic part can contain derivatives with respect to the group variables, which are needed to obtain renormalisable models \cite{BenGeloun:2011rc}, as well as with respect to the scalar fields. Hence its structure is relatively similar to that of standard quantum field theory, even though the interpretation of GFT is very different. 
The kinetic term encodes the propagator of the theory, which can be interpreted as a gluing or identification of lower-dimensional building blocks, here usually pictured as tetrahedra.
In general we can assume an action of the form\footnote{As a technical subtlety, note that the  modes $\varphi_J(\chi^A)$ in (\ref{eq:modeDecomp}) are not real-valued but subject to reality conditions. A simple linear basis change leads to a set of real and independent modes \cite{Calcinari:2024pek}, which are the ones appearing in (\ref{eq:GFTaction}).} 
\be
S[\varphi]= \int {\rm d}^4 \chi \;\mathcal{L},  \qquad \mathcal{L} = \sum_J\left(\frac{1}{2}\mathcal{K}_J^{(0)}\varphi_J^2 - \frac{1}{2}\mathcal{K}_J^{(2)}(\partial_A\varphi_J)^2\right) - V(\varphi)\,,
\label{eq:GFTaction}
\ee
which specifically satisfies the shift symmetry $\chi^A \to \chi^A + \epsilon^A$.
Here, $\pd_A = \frac{\pd}{\pd \chi^A}$ denotes a derivative with respect to scalar field arguments and $\mkz_J$ and $\mkt_J$ are mode-dependent constants. 
$V(\varphi)$ contains all higher-order terms and is often of a complicated nonlocal form; in the geometric interpretation these terms are responsible for generating four-dimensional spacetime histories of the desired combinatorial structure. In applications to cosmology, interactions are often neglected since they are expected to be subdominant in the very early universe (see, e.g., \cite{GFTscalarcosmo,Oriti:2016ueo}), and we will do the same in the following. The notation $\mathcal{K}_J^{(0)}$ and $\mathcal{K}_J^{(2)}$  is taken from \cite{GFTscalarcosmo,Oriti:2016ueo} where these are thought of as expansion coefficients of a derivative expansion that could in principle also include higher-order terms, which we assume are not present. Note that any additional terms in the action would be required to satisfy the shift symmetry with respect to the scalar fields.

While GFT was originally formulated in terms of a functional integral (\ref{eq:GFTZ}), applications to cosmology usually start from a canonical quantisation or more general Hilbert space structure, in which the extraction of effective dynamical equations and study of semiclassical states are more straightforward. This is somewhat similar to LQC which is derived from the canonical, not the covariant approach to LQG. A review of different Hilbert space formalisms for GFT and their foundations can be found in \cite{Gielen_hilber}; we will work in the ``deparametrised'' approach proposed in \cite{GFThamiltonian}, which is essentially a conventional canonical quantisation.
In this setting, after a Legendre transform of \eqref{eq:GFTaction} and a Fourier decomposition with respect to the ``spatial fields'' $\chi^a$ with $a=1,2,3$, one finds that the Hamiltonian of the theory is\footnote{As pointed out in \cite{Gielen:2020fgi}, singling out a clock field, as is necessary to perform the Fourier transform, breaks the initial rotational symmetry between the four fields. }  \cite{GFThamiltonian, Gielen:2020fgi}
\begin{align}
    H = & \piInt{k} \sum_J \frac{\mkt_J}{2}\left( -\frac{1}{|\mkt_J|^2} \pi_{J, -k}\chiz \pi_{J,k} \chiz + \omega_{J,k}^2\,\varphi_{J, -k}\chiz\varphi_{J,k}\chiz \right)\,,
    \label{eq:hamiltonian}
\end{align}
where   $\pi_J = - \mk_J^{(2)} \pd_0 \varphi_J$ is the canonical momentum and we defined $\omega_{J,k}^2 = m_J^2 + \Vec{k}^2$ with $ m_J^2 = - \frac{\mk_J^{(0)}}{\mk_J^{(2)}}$. Note that $m_J^2$ and $\omega_{J, k}^2$ can be negative, depending on the signs of $\mkz_J$ and $\mkt_J$, which depend on the choice of GFT model. Again, being as general as possible, we also include the case $m_J^2<0$ for which $\omega^2_{J,k}<0$ at least for small enough $|\vec{k}|.$

We proceed by promoting the Fourier modes of $\varphi_J$ and its conjugate momentum $\pi_J$ to operators satisfying equal-time commutation relations
\be
[\varphi_{J,k}(\chi^0),\pi_{J',k'}(\chi^0)] = \ii\,\delta_{J J'} (2\pi)^3  \delta(\Vec{k} + \Vec{k'})\,.
\label{eq:commVarphiPi}
\ee
We can then define convenient linear combinations $A_{J, k}, \, A_{J, k}^\dagger$ by
\begin{align}
    \pi_{J,k}\chiz = -{\rm i}\alpha_{J,k}(A_{J,k} - A_{J,-k}^\dagger)\,, \;\; \varphi_{J,k}\chiz = 
    \frac{1}{2\alpha_{J,k}} (A_{J,k} + A_{J,-k}^\dagger)\,; \quad  \alpha_{J,k} = \sqrt{\frac{|\omega_{J,k}| |\mk_J^{(2)}|}{2}}\,,
     \label{eq:piPhiAaDagger} 
\end{align}
which satisfy
\begin{align}
   [A_{J,k}\chiz, A_{J',k'}^\dagger\chiz] = \delta_{J J'} (2\pi)^3 \delta(\Vec{k} - \Vec{k'})\,, 
   \label{eq:commRelTimeDep}
\end{align}
with all other commutators vanishing. Evaluating these operators at time zero defines a set of time-independent creation and annihilation operators by $a_{J,k}=A_{J,k}(0)$ and $a^\dagger_{J,k}=A^\dagger_{J,k}(0)$. 
We find two different types of modes, namely  oscillating and squeezed modes, from the Hamiltonian \eqref{eq:hamiltonian}, depending on the sign of $\omega_{J, k}^2$.
For modes with $\omega_{J,k}^2<0$ the Hamiltonian is a standard harmonic oscillator
\begin{align}
    H_{J,k} = &  -\sgn{\mkt_J} \frac{|\omega_{J,k}|}{2}  \left(a_{J,-k} a_{J,-k}^\dagger + a^\dagger_{J,k} a_{J,k} \right)\,,
    \label{eq:Hosc}
\end{align}
  and the Heisenberg equations of motion $\pd_0 A_{J, k} = - \ii\,  [A_{J, k}, H]$ give
\begin{align}
    A_{J,k} = a_{J,k} e^{\ii\, \sgn{\mkt} |\omega_{J,k}| \chi^0}\,, \quad A_{J,k}^\dagger = a_{J,k}^\dagger e^{-\ii\, \sgn{\mkt} |\omega_{J,k}|\chi^0}\,.
    \label{eq:Adynamics}
\end{align} 
On the other hand, for modes with $\omega_{J,k}^2 >0$ we obtain a squeezing Hamiltonian
\be
    H_{J,k} = \sgn{\mkt_J} \frac{|\omega_{J,k}|}{2}  \left(a_{J,k} a_{J,-k} + a^\dagger_{J,k} a^\dagger_{J,-k} \right)
    \label{eq:Hsq}
\ee
and the time-dependent expressions for our basic operators are given by
\begin{align}
    A_{J,k} = &\, a_{J,k} \cosh \left(|\omega _{J,k}| \chi^0\right)-\ii\, \text{sgn}(\mkt) a_{J,-k}^{\dagger } \sinh \left(|\omega _{J,k}| \chi^0\right)\,,\cr
    A_{J,k}^\dagger= &\, a_{J,k}^\dagger \cosh \left(|\omega _{J,k}| \chi^0\right)+\ii\, \text{sgn}(\mkt) a_{J,-k} \sinh \left(|\omega _{J,k}| \chi^0\right)\,.
    \label{eq:aadSolutions}
\end{align}
Knowledge of these solutions is sufficient to show that the expectation value of the number operator $A^\dagger_{J,k}A_{J,k}$ of squeezed modes satisfies a ``Friedmann equation'' which is asymptotically equivalent to the one of general relativity with a single massless scalar field while resolving the classical singularity, in the sense that only very special initial states can ever have vanishing particle number \cite{Adjei:2017bfm,deparamcosmo}. Interpreting GFT quanta in a fixed mode $J$ as representing spin-network excitations of LQG, a definition of the volume operator similar to the one of LQG would suggest that the total volume is proportional to the number of quanta, and hence a similar Friedmann equation can be obtained for the volume.
In the following analysis we will use the effective metric approach \cite{gftmetric} in which the volume of the universe is a function of this effective metric rather than determined by the number operator, so that the effective Friedmann equation can be different.

For a wide class of possible choices of coefficients $\mk_J^{(0)}$ and $\mk_J^{(2)}$ in the GFT action, including the particularly well-motivated case of a Laplacian operator acting on all group arguments, a single $J$ mode will dominate at late times where the semiclassical limit can be related to general relativity \cite{Gielen:2016uft}. 
Because of this reason and for technical simplicity, we therefore restrict to the analysis of a single Peter--Weyl mode with $J=J_0$ as is common in cosmological GFT studies  \cite{deparamcosmo, Marchetti:2021gcv}. 
If the dominant mode is of squeezed type (as it is when squeezed modes are present at all), the emergence of Friedmann dynamics compatible with general relativity is hence a very general result of GFT cosmology.
An extension of the analysis to multiple modes is straightforward in principle. 
Notice again that 
either the single-mode truncation or a general multi-mode analysis would be possible for many choices of gauge group other than ${\rm SU}(2)$, which makes these results even more general and less sensitive to the details of the GFT model. The statement that such a wide range of GFT models agrees with general relativity may appear ``too good to be true'' from a conceptual point of view, but it only applies to homogeneous and isotropic flat cosmology. Classically,  the correct Friedmann equations can be obtained even from Newtonian dynamics \cite{Mukhanov}, and hence the agreement does not mean that all such models reduce to general relativity at low energies. This  strongly motivates going beyond purely homogeneous spacetimes and including inhomogeneities, where one would expect a possible agreement with general relativity to be much more sensitive to the details of the GFT, as our later results will indeed indicate. 

In the case of a non-compact gauge group (which we will not investigate here), restriction to a single mode is not strictly possible for continuous representations, since these modes are not normalisable. One could choose a sharply peaked Gaussian and obtain qualitatively similar results, as suggested by \cite{Jercher:2021bie}.

Oscillating modes do not lead to an expanding background cosmology and are often not considered, but in a multi-mode analysis only a single squeezed mode is required to obtain an expanding background. 
At the level of perturbations, oscillating modes might then become relevant, at least for a certain time period and depending on initial conditions. Therefore we include them in our later analysis.

\section{Effective GFT metric for a perturbed FLRW spacetime}
\label{sec:effMetric_FLRW}

In \cite{gftmetric} we presented a new proposal for defining an effective spacetime metric in GFT, using symmetries of the GFT action and their relation to symmetries of spacetime fields. We will briefly review this construction in \refsec \ref{sec:GFTmetric_construction}, before considering its specific application to the scenario of an FLRW metric with small inhomogeneities in \refsec \ref{sec:GFTuniverse_effMetric}. 
The application to the FLRW background was already presented in \cite{gftmetric} and our focus is on extending the analysis to perturbations.

\subsection{General construction}
\label{sec:GFTmetric_construction}

The construction of the effective metric relies on using four massless scalar fields as a coordinate system. Such matter reference frames have long been considered in the quantum gravity literature \cite{kuchar_fourFields, Giesel_2019} and were previously employed to study perturbations within GFT \cite{Gielen:2017eco,Gielen_2018, Marchetti:2021gcv, Jercher:2023nxa}.
While some GFT studies \cite{Marchetti:2021gcv, Jercher:2023nxa} include a fifth scalar field that is assumed to dominate the four reference or coordinate fields, so that the matter content can often be approximated as just a single field, we assume that the reference fields constitute the only matter content.
Within classical general relativity, our matter action reads
\begin{align}
S_\chi = \int {\rm d}^4 x\;\mathcal{L}_\chi = -\frac{1}{2}\int {\rm d}^4 x\;\sum_A\, \sqrt{-g}\,  g^{\mu\nu} \partial_\mu\chi^A \partial_\nu\chi^A\,.
\label{eq:GFT_scalaractionFourFields}
\end{align}
Using these fields as coordinates means that we identify each $\chi^A$ with a spacetime coordinate $x^\mu$ by demanding that hypersurfaces of constant $\chi^A$ coincide with hypersurfaces where the respective coordinate is constant. We then have $\partial_\mu\chi^A = \delta^A_\mu$, where $A = 0, 1, 2, 3$ denotes a \emph{label} of the fields (and is not a spacetime index). As before we use $a = 1, 2, 3$ to denote the spatial fields and $0$ for the clock field. For such a relational coordinate system to be locally well-defined, the fields have to satisfy a non-degeneracy condition with respect to any other well-defined coordinate system,\footnote{In the case of only a single scalar field $\chi$ used as a clock, the equivalent condition is $\partial_0\chi\neq 0$: the clock field is not allowed to turn around and evolve backwards.} 
\be
\det(\partial_\mu\chi^A)\neq 0\,.
\label{eq:nondeg}
\ee

The relational coordinate system defines a special case of the harmonic gauge $\Box x^\mu=0$ by virtue of the Klein--Gordon equation  $\Box \chi^A=0$ satisfied by each of the fields.
While the harmonic gauge has a residual gauge freedom, fixing the relational coordinate system as we do here fixes the gauge completely.
Using an Arnowitt--Deser--Misner (ADM) decomposition of the metric, we obtain the following relations between the canonical momenta of the scalar fields $\pi_A = \pd \mathcal{L}_\chi / \pd (\pd_0 \chi^A)$ defined from \eqref{eq:GFT_scalaractionFourFields} and the lapse $N$ and shift $N^a$: 
\begin{align}
\pi_0 = \frac{\sqrt{|q|}}{N}\,, \qquad \pi_a = - N^a \frac{\sqrt{|q|}}{N} = - N^a \pi_0 \,.
\label{eq:momentaRelCoords}
\end{align} 
 These expressions show that the lapse and shift are fully determined by the spatial metric $q_{ab}$ together with the scalar field momenta. This can be seen as an explicit gauge-fixing of the general coordinate freedom of general relativity which would permit an arbitrary choice of $N$ and $N^a$.
 
The action \eqref{eq:GFT_scalaractionFourFields} is invariant under constant shifts of each of the fields $\chi^A\mapsto\chi^A+\epsilon$ ($\epsilon \in \mathbb{R}$), which by virtue of Noether's theorem implies the existence of a current $j^\mu$. There is a separate current for each $\chi^A$, which we can label as $(j^\mu)^A$, satisfiying a conservation law 
\begin{align}
\partial_\mu (j^\mu)^A = 0\,,\qquad (j^\mu)^A =-\sqrt{-g}\, g^{\mu\nu} \partial_\nu\chi^A\,.
\label{eq:classicalCurrentGeneralCoords}
\end{align}
 In the relational setup (with $\partial_\mu\chi^A=\delta_\mu^A$) these currents $(j^\mu)^A$ can be interpreted as a symmetric matrix field $j^{AB} = (j^B)^A$, directly related to the metric via 
\begin{align}
 j^{AB} = -\sqrt{-g}\, g^{A B}
 \label{eq:jmetric}\,.
\end{align}
Hence, in a theory of general relativity coupled to four reference scalars, the spacetime metric can be directly recovered from Noether currents associated to shift symmetries in the matter fields. 

The GFT action \eqref{eq:GFTaction} has an equivalent translational symmetry  $\chi^A \to \chi^A + \epsilon$, here appearing as a translational symmetry on the GFT configuration space. Again, this symmetry is associated with a conserved Noether current, namely the GFT energy-momentum tensor $T^{AB}$ defined by
\be
T^{AB} = -\frac{\partial\mathcal{L}}{\partial (\partial_A\varphi)}\partial_B\varphi+\delta^{AB}\,\mathcal{L} = \sum_J\left(\mathcal{K}_J^{(2)}\partial_A\varphi_J\,\partial_B\varphi_J\right)+\delta^{AB}\,\mathcal{L}\,.
\label{eq:TGFTdef}
\ee
The components of $T^{AB}$ can be promoted to operators and satisfy the conservation law  $\partial_A T^{AB}=0$, both classically and quantum-mechanically, as shown in \cite{gftmetric}. This agrees with the expected classical conservation law $\partial_A j^{AB}=0$ for spacetime Noether currents. The proposal of \cite{gftmetric} is to identify the conserved quantities that arise from the same symmetries in GFT with those of spacetime physics: we view $T^{AB}$ as the GFT version of the classical current $j^{AB}$. (We emphasise that this type of identification holds only in the relational coordinate system.) Hence, $T^{AB}$ encodes the spacetime metric in GFT  through \eqref{eq:jmetric}. The conservation law for $T^{AB}$ translates into the conservation law of $j^{AB}$, which is equivalent to the Klein--Gordon equation for the fields $\chi^A$.
We denote the quantum version of the GFT energy-momentum tensor as $\T^{AB}$ and for sufficiently semiclassical states (such that an interpretation in terms of an effective macroscopic spacetime is justified) we then propose the identification $j^{AB} = \xi \, \expv{\T^{AB}}$, where $ \xi \in \mathbb{R}$, to obtain an effective spacetime metric. The constant $ \xi$, which is not fixed by the general argument based on conservation laws, can be chosen to simplify some expressions and we will set $ \xi=\sgn{\mkt}$.
Note that the metric $g^{AB}$ is not directly represented as an operator in GFT, but only emerges after taking expectation values. In the following analysis, the role of a semiclassical state will be played by the usual Fock coherent state or ``condensate'' used in GFT cosmology.

The quantised GFT energy-momentum tensor $\T^{AB}$ is a function of the annihilation and creation operators $A_{k}, \, A_{k}^\dagger$ (where we have dropped the $J$ label as we focus on a single $J$ mode case in what follows) and explicitly reads 
\begin{align}
\begin{split}
    :\T^{00}_k\!: \; = & \piInt{\gamma}  \frac{\text{sgn}(\mkt)}{4 \sqrt{|\omega_\gamma||\omega_{k-\gamma}|}}\Bigg[ 2 \beta^+_{k, \, \gamma}  :A^{\dagger }_{-\gamma }A_{k-\gamma }: 
    + \beta^-_{k,\,\gamma} \left( :A^{\dagger }_{-\gamma }A^{\dagger }_{\gamma -k}: + :A_{\gamma }A_{k-\gamma }:
   \right)
   \Bigg]\,,\\
   :\T^{0b}_k\!: \; = & \piInt{\gamma} \frac{1}{2}\sqrt{\frac{|\omega_{k-\gamma}|}{|\omega_\gamma|}} \gamma_b
   \left(  :A^{\dagger }_{\gamma
   -k}A_{\gamma }: - :A_{-\gamma }^{\dagger }A_{k-\gamma }:
   - :A_{k-\gamma }A_{\gamma }:
   + :A^{\dagger }_{\gamma -k}A^{\dagger }_{-\gamma }:
   \right)\,,\\
   :\T^{a\neq b}_k\!: \; = & \piInt{\gamma}  \frac{\text{sgn}(\mkt)}{2 \sqrt{|\omega_\gamma||\omega_{k-\gamma}|}}\gamma _a \left(\gamma _b-k_b\right)
   \Big( :A^{\dagger }_{-\gamma }A_{k-\gamma }: + :A_{\gamma -k}^{\dagger }A_{\gamma }: \\
   & \qquad  + :A_{-\gamma }^{\dagger }A_{\gamma -k}^{\dagger }: + :A_{\gamma }A_{k-\gamma }:\Big)\,,\\
   :\T^{aa}_k\!: \; = &  \piInt{\gamma} \frac{\text{sgn}(\mkt)}{4 \sqrt{|\omega_\gamma||\omega_{k-\gamma}|}} \Bigg[ 2(\beta_{k ,\gamma}^- - 2 \gamma_a(k_a - \gamma_a)) :A^{\dagger }_{-\gamma }A_{k-\gamma }: \cr
   & \qquad + (\beta_{k ,\gamma}^+ - 2 \gamma_a(k_a - \gamma_a))\left(:A^{\dagger }_{-\gamma }A^{\dagger }_{\gamma -k}: + :A_{\gamma }A_{k-\gamma }: \right)  \Bigg]\,,
\end{split}
\label{eq:TABoperatorsNormal}
\end{align}
where $\beta^\pm_{k,\gamma}=-m^2+\vec\gamma\cdot(\vec{k}-\vec\gamma)\pm|\omega_\gamma||\omega_{k-\gamma}|$. In these expressions we employ a normal-ordering procedure at the level of the time-independent operators $a_{k},\, a_{k}^\dagger$, such that vacuum expectation values are finite. 
We drop the normal ordering symbol  for the GFT energy-momentum tensor operators in the following, and it should be understood that we always use the normal ordered version of the operators, such that $\expv{\T^{AB}_k}$ stands for $\expv{\no{\T^{AB}_k}}$. 

In summary, the idea is to identify the expectation values of components of the quantum GFT energy-momentum tensor with the classically conserved current, i.e., we propose that we can identify $\expv{\T^{AB}} = \sgn\mkt j^{AB}$ for suitable semiclassical states. One can then reconstruct an effective metric from \eqref{eq:jmetric}. 
In the following we apply this to a perturbed flat FLRW spacetime.

\subsection{Application to an FLRW metric with small inhomogeneities}
\label{sec:GFTuniverse_effMetric}

While the proposal of an effective spacetime metric in GFT is completely general, in this paper we want to specialise to the case most relevant in cosmology, namely a perturbed FLRW metric. This means that we need to calculate the components of the symmetric tensor $j^{AB}$ \eqref{eq:jmetric} resulting from the classically conserved currents for a perturbed FLRW metric, remembering that these are defined only in the relational coordinate system spanned by four massless scalar fields. 
Making the identification of these classical currents with GFT expectation values $\expv{\T^{AB}}$ and inverting the expressions then gives us expressions for metric quantities in terms of operator expectation values in GFT. Since this identification is in a sense a definition of a spacetime metric, in this part we do not require any knowledge of the specific form of $\T^{AB}$ or any choice of state and the expressions below are of a kinematical nature. 
The choice of state and the choice of GFT model will then later determine whether our proposal is sensible, considering also the resulting effective equations of motion for the perturbation variables (see \refsec\ref{sec:GFTuniverse_perturbations} for the analysis for a Fock coherent state). 
We should be able to show that our perturbations of FLRW are indeed small in a well-defined sense, and they should satisfy equations related to classical dynamics of perturbations in some form.

Using standard conventions of cosmology \cite{cosmopert}, the general perturbed FLRW metric reads (here and below $i,j=1,2,3$ denote spatial indices)
\begin{align}
\begin{split}
    \text{d}s^2 =&  -N(t)^2(1+ 2 \tilde{\Phi}(t,\Vec{x}))\text{d} t^2  + 2 N(t) a(t) \,\left( \partial_i B(t, \Vec{x}) - B^V_i(t, \Vec{x}) \right) \, \text{d}t\,\text{d}x^i\\
    & + a(t)^2\Big((1-2 \psi(t, \Vec{x})) \delta_{ij} + 2 \partial_i \partial_j E(t, \Vec{x})\\
    & \qquad \qquad \qquad - \left( \pd_i E^V_j(t, \Vec{x}) + \pd_j E^V_i(t, \Vec{x})\right) + 2 E_{ij}^T(t, \Vec{x})\Big)\text{d}x^i\text{d}x^j,
    \label{eq:generalPerturbedMetric}
\end{split}
\end{align}
where $N$ denotes the background lapse, $a$ the scale factor, and we have carried out a decomposition of metric perturbations into scalar ($\psi\,, \tPhi,\, E,\, B$), vector ($B^V_i, \, E^V_i$) and tensor ($E_{ij}^T$) components. 
The vector components have vanishing divergence and the tensor component is transverse and traceless:
\begin{align}
    \delta^{ij}\pd_j B_i^V = 0, \quad \delta^{ij}\pd_j E^V_i = 0, \quad \delta^{ik}\pd_k E_{ij}^T = 0, \quad \delta^{ij}E_{ij}^T = 0\,.
    \label{eq:pert_conditionsVectorTensor}
\end{align}

For the metric \eqref{eq:generalPerturbedMetric} and the matter action for four massless scalar fields \eqref{eq:GFT_scalaractionFourFields}, we can then obtain the classically conserved currents \emph{in the relational coordinate system}, given by (\ref{eq:jmetric})\footnote{In \refapp \ref{sec:kEssence}, we will generalise this to k-essence models for the scalar fields, which affects the form of the currents.}, as  
\begin{align}
\begin{split}
    j^{00} = & \,  \frac{a^3}{N} \left( 1 -\tPhi - 3\psi + \nabla^2 E \right)\,, \qquad 
    j^{0a} =  a^2 ( B^V_a - \pd_a B)\, ,\\
    j^{a\neq b} = &\,  a N \left( 2 \pd_a \pd_b E - \pd_a E^V_b - \pd_b E^V_a + 2 E_{ab}^T \right)\, ,\\
    j^{aa} = & - a N  \left( 1 + \tPhi - \psi + \nabla^2 E - 2 \pd_a^2 E + 2 \pd_a E^V_a - 2 E_{aa}^T\right)\,  \quad (\text{no sum over} \ a).
    \label{eq:classicalCurr}
\end{split}
\end{align}
We have left the lapse function $N$ general, but it should be understood that the identification of the $j^{AB}$ components with the GFT energy-momentum tensor is only possible in the coordinate system spanned by the four scalar fields with  $N = a^3/ \pi_0$ and $N^a = - \pi_a/\pi_0$ from \eqref{eq:momentaRelCoords}, where $\pi_0$ and $\pi_a$ are the momenta of the clock and rod fields, respectively. 
For scalar perturbations the momenta of the spatial fields are given by $\pi_a = - a^2\pd_a B$.

The conserved current \eqref{eq:classicalCurr} for a flat FLRW universe (i.e., taking into account homogeneous background quantities only) thus takes the form
\begin{align}
j^{AB} = \begin{pmatrix} \pi_0 & 0 \cr 0 & -\frac{a^4}{\pi_0}\delta^{ab}
\end{pmatrix}\,,
\label{eq:jAB_bg}
\end{align}
where we recall that $\pi_0 > 0$. 
Notice that the signs of the components are fixed by the Lorentzian signature of  \eqref{eq:generalPerturbedMetric}; in the case of a Euclidean signature, all entries would be positive. 

The $\T^{AB}$ operators constructed in \eqref{eq:TABoperatorsNormal} are defined in terms of Fourier modes of the spatial fields and we therefore relate them to the Fourier modes of $j^{AB}$. For any classically perturbed quantity we have $f(t,\vec{x}) = \Bar{f}(t) + \delta f(t,\vec{x})$. The background quantity is given by $\Bar{f}(t) = \frac{1}{V_0}\int {\rm d}^3 x \,f(t,\vec{x}) = f_{k=0}(t)$, where $V_0$ is the coordinate volume of the universe or of a patch of the universe (``fiducial cell'') used to define it.
Hence, the $\vec{k}=0$ mode determines the homogeneous part, so that we have $\delta f_{k=0}(t)=0$. By (\ref{eq:jAB_bg}), the conjugate momentum of the clock field and scale factor are then determined by the $\vec{k}=0$ mode of the diagonal components of $\ev{\T^{AB}_0}$:
\begin{align}
    \pi_0 = & \,\sgn{\mkt}\ev{\T^{00}_0} \,, \qquad    a^4 =  - \ev{\T^{00}_0}\ev{\T^{aa}_0}\,.
    \label{eq:bgIdentTAB}
\end{align}
If the off-diagonal components of $\expv{\T^{AB}}$ vanish and all spatial diagonal components $\expv{\T^{aa}_0}$ are identical, the effective metric can consistently be interpreted as \emph{flat} FLRW. We will find this to be the case for the state we consider below.

The non-zero $k$-modes correspond to metric perturbations. In general, the $\expv{\T^{AB}_k}$ can include scalar, vector, and tensor modes according to the decomposition  in \eqref{eq:classicalCurr}. For our choice of state (see (\ref{eq:sigmaGauss}) below) we will see that the operator expectation values can consistently be interpreted as containing only scalar perturbations and we hence neglect vector and tensor perturbations in the expressions that follow. A more complete analysis that reveals which types of state choices can give rise also to vector and tensor perturbations is left for future work. (In particular, it would be of significant interest to include tensor modes and compare these with the linearised form of general relativity or its modifications.)

Assuming only scalar perturbations, the identification $j_k^{AB} = \sgn\mkt \expv{\T^{AB}_k}$  together with (\ref{eq:classicalCurr}) written in Fourier space leads to
\begin{align}
\begin{split}
   \sgn{\mkt} \expv{\T^{00}_{k\neq 0}} = & -\frac{a^3}{N}(\tPhi +3 \psi + k^2 E) , \qquad  \sgn{\mkt}\expv{\T^{0a}_{k\neq 0}} =  -\ii a^2 k_a B,\\
    \sgn{\mkt} \expv{\T^{aa}_{k \neq 0}} = & \, a N  \left( - \tPhi + \psi + k^2 E - 2 k_a^2 E \right)\,  \quad (\text{no sum over} \ a)\,,\\
     \sgn{\mkt}\expv{\T^{a\neq b}_{k\neq 0}} = &  - 2 a N k_a k_b E\,,
    \label{eq:matchedT}
\end{split}
\end{align} 
and therefore $\frac{1}{3}\sgn{\mkt}{\rm tr}\expv{\T^{aa}_{k\neq 0}} = \, a N \left(- \tPhi + \psi + \frac{k^2}{3} E\right)$. 
(Here and in the following ${\rm tr}$ represents a trace over the $a$ index, i.e., ${\rm tr}\expv{\T^{aa}_{k\neq 0}}=\delta_{ab}\expv{\T^{ab}_{k\neq 0}}$.)
Inverting the above gives the following expressions for effective scalar perturbations (we choose $a,b$ so that $k_a \neq 0,\, k_b \neq 0$):\footnote{There is a consistency condition on $E$, which could also be obtained from
\begin{equation*}
E = \frac{\sgn{\mkt}}{2aN(k_b^2-k_a^2)}(\expv{\T^{aa}_{k \neq 0}}-\expv{\T^{bb}_{k \neq 0}})\,, \quad a\neq b\,,
\end{equation*}
and the resulting expression might differ from the one obtained in (\ref{eq:pertQsols}). For the state discussed in \refsec \ref{sec:GFTuniverse_background}, the two expressions agree, see \eqref{eq:TsaddExp} for explicit expressions for squeezed modes and \eqref{eq:pertQsolsOsc} for oscillating modes.} 
\begin{align}
\begin{split}
   \sgn{\mkt}\, \tPhi = & - \frac{\expv{\T^{00}_{k \neq 0}} N}{4 a^3} -\frac{{\rm tr}\expv{\T^{aa}_{k \neq 0}}}{4 a N}\,, \qquad
    \sgn{\mkt}\, E =  -\frac{1}{2 a N}\frac{ \expv{\T^{a\neq b}_{k \neq 0}}}{ k_a k_b } \,,\\
    \sgn{\mkt}\, \psi =  & -\frac{\expv{\T^{00}_{k \neq 0}} N}{4 a^3} + \frac{{\rm tr}\expv{\T^{aa}_{k \neq 0}}}{12 a N}  + \frac{k^2}{k_a k_b} \frac{1}{6 a N} \expv{\T^{a\neq b}_{k \neq 0}}\,, \qquad
   \sgn{\mkt} \, B =    \frac{\ii }{a^2}\frac{\expv{\T^{0a}_{k \neq 0}}}{k_a} \,.
     \label{eq:pertQsols}
\end{split}
\end{align}
These relations are independent of the specific GFT and rely only on the identification  of  expectation values of the GFT energy-momentum tensor $\expv{\T^{AB}}$ with components of a perturbed FLRW metric in a relational coordinate system.
 In particular, they do not depend on a specific choice of GFT state, assuming a state that only contains scalar perturbations.\footnote{This assumption means in particular that all six off-diagonal components of $\expv{\T^{AB}_k}$ can be written in terms of two scalar functions, which is again a consistency check for the proposal.} By making a particular choice of state and computing the corresponding effective metric, one can check explicitly whether the state admits an interpretation as a slightly inhomogeneous universe, by verifying that the background quantities represent an FLRW universe and the non-zero $k$ modes represent small perturbations. The choice of state detailed in \refsec \ref{sec:GFTuniverse_background}  and used already in \cite{gftmetric} should be understood as a naive first guess.

This illustrates the nature of the effective GFT metric proposal: in itself, for any suitably semiclassical state an effective metric can be reconstructed; the task is to interpret its form physically. In \refsec \ref{sec:GFTuniverse_background} and \refsec \ref{sec:GFTuniverse_perturbations} we will compare the effective metric to  a perturbed flat FLRW metric, which reflects our \emph{belief} that our state describes said metric in the semiclassical limit. How to interpret a general metric without assuming a classical counterpart from the beginning is less clear.

\section{Classical analysis}
\label{sec:classicalAnalysis}

Following the philosophy of constructing an effective metric in GFT as described above, defining a suitable state gives us explicit expressions for metric perturbations, subject to equations of motion derived from the GFT dynamics. 
In order to be able to compare the resulting equations to our expectations from classical cosmology, in this section we obtain the dynamical equations for classical cosmological perturbations in a relational coordinate system spanned by four massless scalar fields, as well as for a single massless scalar field that serves as a clock. 
We include the details of the general relativistic system for ease of reference, and we will compare the dynamical equations obtained within GFT to the expressions below in \refsec\ref{sec:GFTuniverse_perturbations}.

In classical spacetime physics, the energy-momentum tensor of four massless scalar fields (defined by an action (\ref{eq:GFT_scalaractionFourFields})) is
\be
 \tensor{T}{^\mu_\nu} = \sum_A\left[g^{\mu\alpha}\partial_\alpha \chi^A  \partial_\nu\chi^A - \frac{1}{2}\delta^\mu_\nu(g^{\alpha\beta}\partial_\alpha \chi^A \partial_\beta\chi^A)\right]\,.
 \label{eq:TscalarFields}
\ee
Let us emphasise that his object has nothing to do with the GFT energy-momentum tensor (\ref{eq:TGFTdef}), which is an abstract field-theoretic quantity not defined on any spacetime. In the gauge where the scalar fields are used as coordinates $\partial_\mu\chi^A=\delta_\mu^A$ we find
\be
 \tensor{T}{^\mu_\nu} = \sum_A\left[g^{\mu A}  \delta_\nu^A - \frac{1}{2}\delta^\mu_\nu\,g^{AA}\right]\,.
 \label{eq:scalarFieldTmunu}
\ee
While originating from a different motivation, models which include three massless scalar fields with homogeneous gradients have been investigated within models of solid inflation \cite{Endlich:2012pz}, including a study of perturbations and differences to more standard cases considered within cosmology. 

\subsection{Background}
\label{sec:pert_fourFields}

At the background level of the perturbed FLRW metric \eqref{eq:generalPerturbedMetric}, we obtain the following expressions for the energy density and pressure (no sum over $i$ in the second expression):
\begin{align}
    -\tensor{T}{^0_0} =\rho = \frac{1}{2} \left(\frac{1}{N^2} + \frac{3}{a^2}\right) = \frac{\pi_0^2}{2 a^6} + \frac{3}{2 a^2}\,, \qquad  \tensor{T}{^i_i} = P = \frac{1}{2} \left(\frac{1}{N^2} - \frac{1}{a^2}\right) = \frac{\pi_0^2}{2 a^6} - \frac{1}{2 a^2}\,.
\end{align}
The contribution of the spatial coordinate fields, coming from their nonvanishing gradient energy, appears as an additional term that would be equivalent to negative spatial curvature ($P=-\frac{1}{3}\rho$).\footnote{These background terms agree with the simplest solid inflation model where $F=-X/2$ in the construction of \cite{Endlich:2012pz}.} For certain initial conditions where $\frac{\pi_0^2}{a^4} \gg 1$, the contribution of the spatial fields to the energy density can become negligibly small for a certain period of time, effectively recovering the standard cosmological background scenario with a single massless scalar field.
This limit can be achieved for sufficiently early times, depending on the value of $\pi_0$, but at late times the gradient energy will always dominate. In general, we have an equation of state parameter $w = P/\rho = \frac{1- a^4/\pi_0^2}{1+ 3 a^4/\pi_0^2} \in (-\frac{1}{3}, 1)$, and similarly for the sound speed $c_s^2 = P'/\rho' = \frac{1 - a^4/(3\pi_0^2)}{1 + a^4/\pi_0^2} \in (- \frac{1}{3}, 1)$. 

The resulting first and second Friedmann equations read (with $\kappa=8\pi G$)
\begin{align}
    H^2 & = \left(\frac{a'}{a}\right)^2 = \frac{\kappa}{6} \left( 1 + 3 \frac{a^4}{\pi_0^2} \right), \qquad \frac{a''}{a} = \frac{\kappa}{6} \left( 1 + 9 \frac{a^4}{\pi_0^2} \right),
\label{eq:clFried}
\end{align}
where the terms proportional to $\frac{a^4}{\pi_0^2}$ arise due to the spatial fields and would not appear in the case of a single (clock) scalar field. 
An alternative way of writing the second Friedmann equation is $H' = \kappa \frac{a^4}{\pi_0^2}$. Again it is clear that the contributions from gradient energy will always dominate at late times; the solutions for $a$ will actually diverge as $|\chi^0-\chi_{\rm f}|^{-1/2}$ at some finite value $\chi^0=\chi_{\rm f}$.

\subsection{Perturbations}
\label{sec:classicalAnalysis_perturbations}

In the following, we give the results for the perturbative analysis for the gauge-fixed system with four massless scalar fields.  We are interested in the dynamics of perturbations of a system with four fields to compare with the GFT equations; a more detailed classical analysis of the general-relativistic system is not the focus of the current work. 

The harmonic gauge conditions $\Box x^\mu=0$ for our choice of lapse $N = a^3/\pi_0$ read (these equations agree with the ones given in \cite{Marchetti:2021gcv})
\begin{align}
\begin{split}
  a^2\nabla^2 B + \pi_0(\tPhi + 3 \psi - \nabla^2 E)' = &\, 0 \,,\\
\vec\nabla\left[- \pi_0 (2 H B+ B') + a^2(- \tPhi + \psi + \nabla^2 E)\right] = &\, 0\, .
\label{eq:harmGaugeCond}
\end{split}
\end{align}
These can be used to eliminate $\nabla^2 B$ for example, and combined to give
\begin{align}
\pi_0^2 (- \tPhi - 3 \psi + \nabla^2 E)'' = a^4 \nabla^2(- \tPhi + \psi + \nabla^2 E)\,.
\label{eq:combinedHGC}
\end{align}

The components of the perturbed Einstein tensor $\tensor{\delta G}{^\mu_\nu}$ for our lapse choice are given by 
\begin{align}
    \tensor{\delta G}{^0_0} = &\, \frac{6 \pi_0^2}{ a^6} H \left(H \tPhi +  \psi'\right) +   \frac{2 \pi_0}{a^4} H \nabla^2 B - \frac{2}{a^2}\nabla^2 \psi - \frac{2 \pi_0^2}{ a^6}H\nabla^2 E' \,, \\
    \tensor{\delta G}{^0_i} = &\, - \frac{2 \pi_0^2}{ a^6} \pd_i\left( H \tPhi + \psi' \right)\,,\\
    \begin{split}
    \tensor{\delta G}{^i_i} = &\,  \frac{2 \pi_0^2}{ a^6} \left(2 H' - 3 H^2\right) \tPhi + \frac{2 \pi_0^2}{ a^6} H\tPhi' + \frac{2 \pi_0^2}{ a^6}\psi''\\
    & + (\nabla^2 - \pd_i^2) \left( \frac{\pi_0}{ a^4} \left( 2 H B + B'\right) + \frac{1}{a^2} (\tPhi - \psi )  - \frac{\pi_0^2}{ a^6} E''\right)\,, 
    \end{split}\\
    \tensor{\delta G}{^i_{\neq j}} = &\, \pd_i \pd_j \left( -\frac{\pi_0}{a^4}\left(2 H B + B'  \right) -\frac{1}{a^2}(\tPhi - \psi)  + \frac{\pi_0^2}{ a^6} E'' \right)\,.
  \label{eq:pert_einsteinPerturbed}
\end{align}

Our choice of coordinate system naturally limits us to the harmonic gauge, in fact, it completely fixes the gauge and eliminates the residual gauge freedom. 
In particular, there are no perturbations in the scalar fields in the relational coordinate system where $\pd_\mu \chi^A = \delta_\mu^A$.
For the perturbed energy-momentum tensor for four massless scalar fields in the relational coordinate system we find
\begin{align}
\begin{split}
 \tensor{\delta T}{^0_0} = &  \frac{\pi_0^2}{a^6}\tPhi - \frac{1}{a^2} (3 \psi - \nabla^2 E)\,, \qquad  \tensor{\delta T}{^0_i} =  \frac{\pi_0}{a^4} \pd_i B  = \tensor{\delta T}{^i_0} \,,\\
   \tensor{\delta T}{^i_{\neq j}} = & - \frac{2}{a^2} \pd_i \pd_j E \,, \qquad 
 \tensor{\delta T}{^i_i} =  -\frac{\pi_0^2}{a^6}\tPhi+ \frac{1}{a^2}\left(  -\psi + \left(\nabla^2 - 2 \pd_i^2\right) E \right)\,.
 \label{eq:TchiFourFields}
 \end{split}
\end{align}
Note that, unlike in the single field case \eqref{eq:TchiOneField}, $\tensor{\delta T}{^0_i}\neq 0$ due to the sum over spatial fields in \eqref{eq:scalarFieldTmunu}. 
From the perturbed Einstein equations $\delta \tensor{G}{^\mu_\nu}=\kappa \delta \tensor{T}{^\mu_\nu}$ and the harmonic gauge conditions \eqref{eq:harmGaugeCond} one can derive equations of motion for all metric perturbation variables. First of all, we have
\begin{align}
\begin{split}
0=\tensor{G}{^i_{\neq j}}-\kappa \delta \tensor{T}{^i_{\neq j}}= &\,\partial_i\partial_j\left( -\frac{\pi_0}{a^4}\left(2 H B + B'  \right) -\frac{1}{a^2}(\tPhi - \psi)  + \frac{\pi_0^2}{ a^6} E'' +\frac{2\kappa}{a^2} E \right) \\
= &\,\partial_i\partial_j\left(-\frac{1}{a^2}\nabla^2 E  + \frac{\pi_0^2}{ a^6} E'' +\frac{2\kappa}{a^2} E \right)
\end{split}
\end{align}
and hence
\be
E'' - \frac{a^4}{\pi_0^2}\nabla^2 E +2  \kappa \frac{a^4}{\pi_0^2}  E = 0 \,.
\label{eq:Eeq}
\ee
Then, from $\delta \tensor{G}{^0_0}-\delta \tensor{G}{^i_i}=\kappa \delta \tensor{T}{^0_0}-\kappa\delta \tensor{T}{^i_i}$ (with sum over $i$), elimination of $\nabla^2 B$ from the harmonic gauge condition and use of the background equations (\ref{eq:clFried}), we have
\be
\tPhi'' - 4 H \tPhi' - \frac{a^4}{\pi_0^2}\nabla^2\tPhi =  0\,.
\label{eq:Psieq}
\ee
Furthermore we can obtain $\psi$ and $B$ from the $(0,0)$ and $(0,i)$ components of the Einstein equations:
\begin{align}
  -2 \nabla^2\psi + 3 \kappa   \psi = -3 \kappa   \tPhi + 2\frac{\pi_0^2}{a^4} H\tPhi' +  \kappa \nabla^2 E    \,, \qquad  - \frac{2 \pi_0}{a^2} \left( H \tPhi + \psi' \right) =  \kappa  B\,,
\end{align}
again after eliminating $\nabla^2 B$ and using the background equations to replace $H^2$.
 
 Starting with (\ref{eq:combinedHGC}) and using (\ref{eq:Eeq}), (\ref{eq:Psieq}) and $\delta \tensor{G}{^0_0}=\kappa \delta \tensor{T}{^0_0}$, we can also obtain
 \be
 \psi''-\frac{a^4}{\pi_0^2}\nabla^2\psi+2\kappa\frac{a^4}{\pi_0^2}(\psi+\tPhi)=0\,,
 \ee
 which is almost an equation for $\psi$ alone.

In the following we will also want to compare with the corresponding equations for the case of a \emph{single} massless scalar field $\chi$ used as a clock, $\partial_\mu\chi=\delta_\mu^0$. In this case, the perturbed energy-momentum tensor is
\begin{equation}
 \tensor{\delta T}{^0_0} =  \frac{\pi_0^2}{a^6}\tPhi\,, \qquad  \tensor{\delta T}{^0_i} =  0,\qquad \tensor{\delta T}{^i_0} = \frac{\pi_0}{a^4} \pd_i B\,, \qquad\tensor{\delta T}{^i_{\neq j}} = 0 \,, \qquad 
 \tensor{\delta T}{^i_i} =  -\frac{\pi_0^2}{a^6}\tPhi\,,
 \label{eq:TchiOneField}
\end{equation}
and we find the following equations of motion for the perturbations:
\begin{align}
E'' - \frac{a^4}{\pi_0^2}\nabla^2 E = 0\,, \qquad 
\tPhi'' - 4 H \tPhi' - \frac{a^4}{\pi_0^2 }\nabla^2 \tPhi = 0\,, \qquad \psi'' - \frac{a^4}{\pi_0^2} \nabla^2 \psi = 0\,.
\label{eq:pert_eomSingleFieldpert}
\end{align}
The $\psi$ equation follows from (\ref{eq:combinedHGC}) (which which is a gauge condition and therefore holds independently of the matter content) in combination with the equations for $E$ and $\Phi$, and the $(0,0)$ and $(0,i)$ parts of the Einstein equations with (\ref{eq:TchiOneField}).

In this case of a single field, where the background solution is $a=a_0\exp(\mathcal{H}\chi)$, one can obtain the explicit solution in Fourier space
\begin{align}
\label{eq:Eclassical}
E(\chi) = c_1(k)J_0\left(\frac{a_0^2}{2\mathcal{H}|\pi_0|}|k|e^{2\mathcal{H}\chi}\right) + c_2(k)Y_0\left(\frac{a_0^2}{2\mathcal{H}|\pi_0|}|k|e^{2\mathcal{H}\chi}\right)\,,
\end{align}
where $J_0$ and $Y_0$ are Bessel functions of the first and second kind, respectively, and $c_1(k)$ and $c_2(k)$ are initial condition parameters. These solutions have also been derived in \cite{Marchetti:2021gcv}. These modes oscillate rapidly with growing $\chi$ while their amplitude falls off as $e^{-\mathcal{H}\chi}$, or as $1/a$.

For completeness, we also report the form of two  gauge-invariant variables commonly used in the literature, namely
the curvature perturbation on equal density hypersurfaces $\zeta$ and the comoving curvature perturbation $\mr$. The latter satisfies $\pd_i \mr = \pd_i \psi + H \delta \tensor{T}{^0_i}/(\rho + P)$ and they take on the following form in the relational coordinate system:\footnote{Note that the use of the symbols $\zeta$ and $\mr$ is not consistent across the literature. We use the same convention as, e.g., \cite{PeterUzan, Baumann_book}, but the opposite of \cite{Durrer_book}.} 
\begin{align}
-\zeta  := \psi + \frac{H}{\rho'}\delta\rho = \psi + \frac{1}{3}\frac{\frac{\pi_0^2}{ a^{4}} \tPhi - (3 \psi - \nabla^2 E)}{1 + \frac{\pi_0^2}{ a^{4}} } \,, \qquad \mr = \psi + \frac{ \pi_0}{a^2}\frac{H B}{1 + \frac{\pi_0^2}{ a^{4}}}\,,
\end{align}
where $\delta \rho = - \delta \tensor{T}{^0_0}$ is the perturbed energy density.

As for the background, the above reduces to the single field case in the limit $\frac{\pi_0^2}{ a^{4}} \gg 1$. In particular, in this limit we find $-\zeta \to \psi + \frac{\tPhi}{3}$ and $\mr \to \psi$. 
We do not use these expressions further in this paper given that the dynamics of perturbations will be found to disagree with general relativity, but in future work one can use the effective scalar perturbations obtained as described in \refsec \ref{sec:GFTuniverse_perturbations} to study gauge-invariant quantities from GFT explicitly, which was not possible prior to the proposal of an effective GFT metric.

\section{Emergent FLRW universe from a coherent state}
\label{sec:GFTuniverse_background}

To obtain explicit expressions for the operator expectation values, enabling us to concretely reconstruct an FLRW metric as well as its perturbations from the identifications  \eqref{eq:bgIdentTAB} and \eqref{eq:matchedT}, we have to make a choice of state. 
We use the same state as in \cite{gftmetric}, which was chosen based on the condition of semiclassicality, such that the expectation values $\expv{\T^{AB}}$ can indeed be related to an effective metric, as well as the requirement that it must incorporate properties of the cosmological spacetime. 
Fock coherent states satisfy the requirement of relatively small uncertainty in operator expectation values throughout the evolution  \cite{deparamcosmo}  (see also \cite{Calcinari:2023sax} for a more in-depth analysis of a broader class of semiclassical GFT states). 
We work with a Fock coherent state $\ket{\sigma}$ which is an eigenstate of the (time-independent) annihilation operator $ a_{J,k}\ket{\sigma} = \sigma_J(\Vec{k}) \ket{\sigma}$: 
\begin{align}
   \ket{\sigma} = e^{- ||\sigma||^2/2} \exp\left(\sum_J \int \frac{{\rm d}^3k}{(2\pi)^3} \sigma_J(\Vec{k}) a^\dagger_{J,k}\right)|0\rangle\,,
    \label{eq:sigmaGauss}
\end{align}
where $|0\rangle$ is the GFT Fock vacuum and $||\sigma||^2 = \sum_J \piInt{k} |\sigma_J(\Vec{k})|^2$. 
To reflect the homogeneity of the FLRW metric in the quantum state, we choose a sharply peaked Gaussian for $\sigma(\vec{k})$,
\begin{align}
  \sigma_J(\Vec{k}) = \delta_{J,J_0}\frac{\ma + \ii \mb}{c_\sigma} \ e^{-\frac{(\Vec{k}-\Vec{k}_0)^2}{2 s^2}} \, ,
  \label{eq:sigmaChoice}
\end{align}
where $\ma, \, \mb \in \mathbb{R}$, $s $ determines the peakedness of the state, and we set the homogeneous mode as  the initially dominantly excited Fourier mode, i.e., $\Vec{k}_0 = 0$. 
The normalisation factor $c_\sigma = \left(\frac{s}{2 \sqrt{\pi}}\right)^{3/2}$ is fixed for convenience regarding later calculations. 
The state reflects our restriction to a single Peter--Weyl mode; in the more general case of multiple modes, the initial conditions, namely $\ma, \ \mb$ and $s$, could be  $J-$dependent. 
While the Gaussian is strongly peaked on the background mode, it has a finite width, such that inhomogeneous modes with $\vec{k} \neq 0$ will always be excited. A strictly homogeneous state is reached in the limit of $s \to 0$, corresponding to an infinitely peaked state, which would introduce divergences that are avoided for $0<s\ll 1$. 
In standard cosmological approaches, one treats the perturbations and the background independently, where the background is either classical or a quantum state for a minisuperspace model, and the state for perturbations is defined separately. 
This is conceptually different to our proposed state, which does not allow to excite solely the homogeneous background. 
Since we have chosen $\sigma(\vec{k})$ to be sharply peaked on the background mode $\vec{k}=0$, modes with low $|\vec{k}|$ will have the dominant contribution to the expectation value of the energy-momentum tensor \eqref{eq:TABoperatorsNormal} in addition to the background mode.\footnote{For squeezed modes, this statement no longer holds for (very) large values of $|\chi^0|$, as large $|\vec{k}|$ modes have a larger growth rate $\omega_k$.}

The effective FLRW metric resulting from the homogeneous mode $\vec{k}=0$ was discussed in detail in \cite{gftmetric}; here we include only a brief recap of these results. Inhomogeneous modes are discussed in detail in \refsec \ref{sec:GFTuniverse_perturbations}. 

From the identifications $\sgn{\mkt} \expv{\T^{AB}_{k= 0}} = j_{k= 0}^{AB}$  the explicit form of the effective FLRW metric follows from \eqref{eq:jAB_bg}.   
For our choice of state \eqref{eq:sigmaGauss}, convolutions appearing in the operator expressions \eqref{eq:TABoperatorsNormal} can be simplified with the saddle-point approximation (which we will also use to calculate expressions for the $\vec{k}\neq 0$ modes)
\begin{align}
    \int \dd^3 x \;e^{-\frac{(\Vec{x}-\Vec{\mu} )^2}{s^2}} g(\Vec{x}) \approx  g(\Vec{\mu}) \int \dd^3 x \;e^{-\frac{(\Vec{x}-\Vec{\mu} )^2}{ s^2}} = g(\Vec{\mu}) (\sqrt{\pi} s)^3\,.
    \label{eq:saddlePtApproxLO_text}
\end{align}
This approximation holds for sharply peaked Gaussians such that $g(\vec{x})$ can be considered approximately constant in the region $|\vec{x}-\vec{\mu}|\le s$ and is applicable for our state choice due to $\sigma(\vec{k})$ being highly peaked.
As investigated in \cite{gftmetric} using such an approximation naturally limits the time span for which our analytic expressions are sufficiently accurate. 

The dynamics of $\T^{AB}$ depend on the type of modes we are considering -- squeezed \eqref{eq:aadSolutions} or oscillating \eqref{eq:Adynamics}. 
As we are interested in recovering an expanding universe, our focus lies on squeezed modes, which have a growing number of quanta over time. 
We also report the contribution to an effective metric from oscillating modes for completeness.

Through the identification \eqref{eq:jAB_bg} the signs of the components of the conserved current are directly related to the metric signature: all entries of the conserved current will either have the same sign (Euclidean case) or the spatial diagonal will have the opposite sign of the $j^{00}$ entry (Lorentzian case). 
The initial conditions $\ma, \ \mb$ in \eqref{eq:sigmaChoice} determine the signature of the effective metric we reconstruct;
the Lorentzian case is found for $\mb^2>\ma^2$, whereas $\mb^2<\ma^2$ results in a Euclidean metric.\footnote{The special case of $\mb^2 = \ma^2$ corresponds to vanishing momentum of the clock field and is therefore excluded.} 
Here we are interested in the Lorentzian case and therefore restrict ourselves to initial conditions with $\mb^2>\ma^2$. In connection to earlier discussions, we can note that the effective metric signature is determined by initial conditions in the state rather than any particular features of the underlying GFT model, such as a choice of gauge group. A similar dependence on initial conditions rather than definitions of the GFT model was observed in \cite{Marchetti:2021gcv}.
 
Comparison with the conserved current for the FLRW case as given in \eqref{eq:bgIdentTAB} then gives the following identifications for the momentum of the clock field and the scale factor in the case of squeezed modes $m^2_{J_0} = m^2 >0$  (the expectation values $\ev{\T^{AB}_0}$ follow from the $\vec{k}\rightarrow 0$ limit of our later more general expression (\ref{eq:TsaddExp}))
\begin{align}
\begin{split}
    \pi_0 = & \sgn{\mkt} \ev{\T^{00}_0} =  |m| (\mb^2 - \ma^2), \\
   a^4 = & - \sgn{\mkt} \pi_0  \ev{\T^{aa}_0} =    m^2 (\mb^2 - \ma^2)\left( (\ma^2 + \mb^2)\cosh(2 |m| \chi^0) - 2 \sgn{\mkt} \ma \mb \sinh(2 |m| \chi^0) \right)\\
   = &     \frac{m^2}{2}(\mb^2 - \ma^2)\left( (\ma - \sgn{\mkt} \mb)^2 e^{2 |m| \chi^0} + (\ma + \sgn{\mkt} \mb)^2 e^{- 2 |m| \chi^0} \right)  \,.
\end{split}
\label{eq:jTident}
\end{align}
Importantly, the off-diagonal components $\expv{\T^{0a}_0}$ and $\expv{\T^{a\neq b}_0}$ vanish exactly due to the antisymmetry of the integrals, giving a spatially flat metric.

From the above we obtain the following effective Friedmann equation: 
\begin{align}
\begin{split}
    H^2  = \left( \frac{a'}{a}\right)^2  
    = \frac{1}{4} m^2 \left(1 - \frac{4(\ma^2 - \mb^2)^2}{((\ma - \sgn{\mkt} \mb)^2 e^{2 m \chi^0} + (\ma + \sgn{\mkt} \mb)^2 e^{-2 m \chi^0})^2} \right) 
     = &  \frac{1}{4} m^2 \left(1 - \frac{\pi_0^4}{a^8} \right)\\ 
      \quad
     \underset{\rm late \ times}{\longrightarrow} & \quad \frac{1}{4} m^2.
\label{eq:effFried}
\end{split}
\end{align}
In addition to a constant Hubble rate at late times, the effective metric gives a bouncing universe, with the bounce occurring at $a^4=\pi_0^2$, or equivalently, $\ev{\T^{aa}_0}^2 = \ev{\T^{00}_0}^2$. 
The Ricci scalar at the bounce reads $R_{\rm bounce} = 6 \frac{m^2}{\pi_0}$, thus resolving the singularity of the classical scenario. Singularity resolution through a bounce is a common feature of GFT cosmology models with a single scalar matter field (see, e.g., \cite{GFTscalarcosmo,Oriti:2016ueo,Adjei:2017bfm,deparamcosmo}), but in these past works the Hubble rate is derived from the time evolution of a total volume proportional to the number operator, which is different from our proposal using the effective GFT metric. 
Recovering a constant Hubble rate in the late-time limit is in agreement with all the Friedmann equations previously obtained for GFT models as well as with the general relativistic Friedmann equation for a single massless scalar field if we fix $m^2 = \frac{2}{3}\kappa$. However, the Friedmann equation in general relativity with four massless scalar fields in \eqref{eq:clFried} is different, since the gradients of the spatial fields contribute. Hence, there is a mismatch with what one might expect from the underlying cosmological model already at the background level.
This mismatch is discussed already in the introduction of this paper as well as in \cite{gftmetric}, which focused on the background dynamics.
Another difference with past GFT work (and other scenarios such as loop quantum cosmology) is that the bounce is not associated to a maximal value of the energy density in the scalar field, but can occur at either high or low curvatures. Indeed, the Ricci scalar at the bounce depends on the initial condition set by $\pi_0$.

In the case of oscillating modes,  $\sigma(\vec{k})$ needs to be especially peaked, so that contributions from squeezing modes can be neglected in the integral and only the region near $\vec\gamma=0$ (which consists entirely of oscillating modes) contributes. We then obtain 
\begin{align}
\begin{split}    
\pi_0  = &\; |m| (\ma^2 + \mb^2)\, ,\\
    a^4 = & -|m|^2  (\ma^2 + \mb^2)\left( (\ma^2 - \mb^2) \cos(2 |m| \chi^0) - 2 \,\sgn{\mkt} \ma \mb \sin(2 |m| \chi^0)\right)\,.
\end{split}
\end{align}
The sign of $\pi_0$ is independent of the initial conditions.
The sign of $a^4$ is not fixed and fluctuates throughout the evolution, such that a single oscillating mode would lead to a metric with variable signature; see the discussion above \eqref{eq:jTident}. 
Phenomenologically, oscillating modes can introduce a possible modulation to the evolution of the background scale factor if they appear in conjunction with at least one squeezed mode. 

This concludes the discussion of the cosmological background metric, as reconstructed from the $\vec{k}=0$ mode of the GFT energy-momentum tensor for a suitable state. 
For a squeezed Peter--Weyl mode, we recover an effective expression for the scale factor that leads to an effective Friedmann equation with a bounce. In the following we will extend the analysis to inhomogeneous modes.

\section{Cosmological perturbations}
\label{sec:GFTuniverse_perturbations}

We now focus on the $\vec{k}\neq 0 $ modes of the GFT energy-momentum tensor $\T^{AB}$ \eqref{eq:TABoperatorsNormal} for the state \eqref{eq:sigmaChoice} introduced in \refsec \ref{sec:GFTuniverse_background}. Recall that even though the state is highly peaked on the homogeneous mode, inhomogeneous modes will always be excited. 
In the following we examine the dynamics that arise for cosmological perturbations if we identify these inhomogeneous modes with components of the perturbed FLRW metric \eqref{eq:generalPerturbedMetric}.
Perhaps unsurprisingly given that we are working in a simple approximation to the full GFT and with the simplest possible state, we find a mismatch with the dynamics of general relativity. 
Still, the following can be seen as a guidance to construct perturbative quantities and may give hints which adjustments could lead to an agreement with general relativity at late times. 

All components of the GFT energy-momentum tensor \eqref{eq:TABoperatorsNormal} depend on the same operator combinations; 
in particular, each term is a product of time-dependent ladder operators $A_k$ and $A_k^\dagger$. From the state choice \eqref{eq:sigmaGauss} with \eqref{eq:sigmaChoice} and the linear dependence of $A_k\, , A_k^\dagger$ on the time-independent creation and annihilation operators (see \eqref{eq:Adynamics} and \eqref{eq:aadSolutions}) we find that each of the terms in the expectation values for $\expv{\T^{AB}_k}$ will be proportional to $e^{-\frac{\vec\gamma^2}{2 s^2}} e^{- \frac{(\vec{k}-\vec\gamma)^2}{2 s^2}}$. 
Similarly to the background dynamics, we can then employ the saddle-point approximation \eqref{eq:saddlePtApproxLO_text} to obtain explicit dynamics for the $\expv{\T^{AB}}$ components. 
For this, we rewrite the exponentials appearing in the integrals as
\begin{align}
     e^{-\frac{\vec{k}^2-2 \vec{k}\cdot \vec\gamma+2 \vec\gamma^2}{2 s^2}} = e^{-\frac{1}{s^2}(\vec\gamma-\frac{\vec{k}}{2})^2}e^{-\frac{\vec{k}^2}{4s^2}},
\end{align}
so that the saddle-point approximation implies  $\vec{\gamma}\approx \frac{\vec{k}}{2}$. 
This approximation, which requires $s\ll 1$, will not hold for all times or for large values of $k$. 
Note furthermore that for our choice of $\sigma(\vec{k})$ \eqref{eq:sigmaChoice} we have $A_k \ket{\sigma} = A_{-k}\ket{\sigma}$ (and similarly for $A_k^\dagger$) 
  for oscillating as well as squeezed modes, due to $\omega_k = \omega_{-k}$. For squeezed modes the operator expectation values \eqref{eq:TABoperatorsNormal} then simplify to 
\begin{align}
\begin{split}
    \expv{\T^{00}_k} \; \approx &\,    \frac{\text{sgn}(\mkt)}{4 |\omega_{k/2}|} c_\sigma^2 \Bigg[ k^2\expv{ :A^{\dagger }_{k/2 }A_{k/2 }:} 
    -2 m^2 \left( \expv{:A^{\dagger }_{k/2 }\!^2:} + \expv{:A_{k/2 }\!^2:}
   \right)
   \Bigg]\,,\\
   \expv{\T^{0b}_k} \; \approx & \, \frac{k_b}{4} 
   c_\sigma^2\Bigg[    \expv{:A^{\dagger }_{k/2} \!^2:} 
   - \expv{:A_{k/2 }\!^2:}
   \Bigg]\,,\\
   \expv{\T^{a\neq b}_k} \; \approx &\,  - \frac{\text{sgn}(\mkt)}{ |\omega_{k/2}|} \frac{k_a k_b}{8}
   c_\sigma^2\Bigg[2 \expv{:A^{\dagger }_{k/2 }A_{k/2 }:}  + \expv{:A_{k/2 }^{\dagger }\!^2:} +  \expv{:A_{k/2 }\!^2:}\Bigg]\,,\\
   \expv{\T^{aa}_k} \; \approx & \,  \frac{\text{sgn}(\mkt)}{4 |\omega_{k/2}|} c_\sigma^2\Bigg[ -(4 m^2 + k_a^2 )\expv{ :A^{\dagger }_{k/2 }A_{k/2 }:} 
   + \frac{k^2 -  k_a^2}{2}\left(\expv{:A^{\dagger }_{k/2 }\!^2:} + \expv{:A_{k/2 }\!^2:} \right)  \Bigg]\,,\
\end{split}
\label{eq:TABoperatorsNormal_saddPt}
\end{align}
where the factor $c_\sigma^2$ enters from the integral over the exponential in the saddle-point approximation \eqref{eq:saddlePtApproxLO_text} and is cancelled by our choice of state \eqref{eq:sigmaChoice} in later expressions. 
We use equality signs in the expressions that follow; it should be understood that statements below rely on the applicability and sufficient accuracy of the saddle-point approximation.

For oscillating modes we have different expressions of $\beta^\pm$ in \eqref{eq:TABoperatorsNormal} and therefore the following components differ from the squeezed case:
\begin{align}
\begin{split}
    \expv{\T^{00}_k} \; \approx &\,    \frac{\text{sgn}(\mkt)}{4 |\omega_{k/2}|} c_\sigma^2 \Bigg[ 4 |m^2|\expv{ :A^{\dagger }_{k/2 }A_{k/2 }:} 
    + \frac{k^2}{2} \left( \expv{:A^{\dagger }_{k/2 }\!^2:} + \expv{:A_{k/2 }\!^2:}
   \right)
   \Bigg]\,,\\
   \expv{\T^{aa}_k} \; \approx & \,  \frac{\text{sgn}(\mkt)}{4 |\omega_{k/2}|} c_\sigma^2\Bigg[ \left(k^2 - k_a^2 \right)\expv{ :A^{\dagger }_{k/2 }A_{k/2 }:} 
   + \left(2|m^2| -  \frac{k_a^2}{2}\right)\left(\expv{:A^{\dagger }_{k/2 }\!^2:} + \expv{:A_{k/2 }\!^2:} \right)  \Bigg]\,.
\end{split}
\label{eq:TABoperatorsNormal_saddPt_osc}
\end{align}

As detailed in the previous section, recovering a Lorentzian or Euclidean FLRW background metric with a single Peter--Weyl mode is only possible in the case of a squeezed mode. 
Since there is no split between background and perturbations in our formalism, perturbations are then also of squeezing type in the single-mode case, $J = J_0$. 
In the more general case, where a minimum of two $J$ modes are excited, one of them can be of the oscillating type, as this will not alter the background dynamics at late times. For completeness we then also consider the perturbations arising from oscillating modes.\\

From the relation of perturbation variables to operator expectation values as given in \eqref{eq:pertQsols} we can establish equations of motion for effective perturbations arising from the GFT effective metric in terms of the dynamics of operator expectation values independent of the explicit state choice.
From the identifications in \eqref{eq:pertQsols} we obtain  
the following equations of motion for $E$ and $B$, as well as for the combination $\tPhi - \psi$, which gives a particularly convenient form,
\begin{align}
\begin{split}
   B'' + 4 H B' + 2 \left(  H' +2  H^2  \right) B = & \ \ii\,  \sgn{\mkt} \, \frac{\expv{\T^{0a}}''}{k_a a^2}\,,\\
E''+ 8 H E' + 4 \left(H' + 4 H^2 \right) E = &   - \frac{\sgn{\mkt} \pi_0 }{2 k_a k_b a^4} \expv{\T^{a\neq b}}''\, ,
\label{eq:generalEdynamicsTAB}\\
    (\tPhi - \psi)'' + 8 H  (\tPhi ' - \psi') +4  \left(H' +4 H ^2\right) (\tPhi - \psi) = & 
      - \frac{\sgn{\mkt} \pi_0}{6 k_a k_b  a ^4}(2 k_a k_b{\rm tr}
   \expv{\T^{aa}}'' +k^2 \expv{\T^{a\neq b}}'')\,.\\
   \end{split}
\end{align}
We proceed to analyse squeezed and oscillating modes separately, 
 due to their differing late time limits, where 
we explicitly compute the expressions of effective scalar perturbations for squeezed and oscillating modes in \refsec\ref{sec:squeezedPert} and \ref{sec:oscPert}, respectively. 
The classical analysis for four massless scalar fields and a single field was carried out in \refsec\ref{sec:classicalAnalysis}.
We will focus on comparing the dynamics of the scalar perturbation $E$ as obtained from the quantum theory  to those of general relativity, due to its comparative simplicity. 
As the effective Friedmann equation derived in \eqref{eq:effFried} has the late time limit of general relativity with a single scalar field without a contribution from spatial gradients, we compare the effective GFT perturbation equations to the single field case as well.  
In principle, one could carry out a comparative analysis for all scalar perturbation variables, however, 
as we will find a considerable mismatch between effective GFT dynamics and general relativity, focusing on $E$ 
 should suffice at this stage. The full analysis would presumably become more relevant once agreement with general relativity has been established in the late-time regime.

\subsection{Squeezed modes}
\label{sec:squeezedPert}

The inhomogeneous squeezed modes, which we recall have $\omega_k^2> 0$, have similar dynamics to the background mode with additional $k-$dependent terms. In particular, all components of $\expv{\T^{AB}_k}$ grow exponentially. 
To obtain explicitly their dynamics from \eqref{eq:TABoperatorsNormal_saddPt} it is useful to define the following expressions 
\begin{align}
\begin{split}
 \expv{ :A^{\dagger }_{k/2 }A_{k/2 }:} & =   \, \frac{e^{-\frac{k^2}{4s^2}}}{2 c_\sigma^2}  \left( (\ma - \sgn{\mkt}\mb)^2 e^{2 |\omega_{k/2}| \chi^0} + (\ma + \sgn{\mkt}\mb)^2 e^{-2 |\omega_{k/2}| \chi^0} \right)\\
 & =: \, \frac{1}{c_\sigma^2} \nk \,, \\
 \expv{:{A^{\dagger }_{k/2 }}^2:} + \expv{:{A_{k/2 }}^2:}  & =  \,  \frac{2}{c_\sigma^2}\,  e^{-\frac{k^2}{4s^2}} (\ma^2 - \mb^2) =: \frac{1}{c_\sigma^2} \ck \,,
 \label{eq:defNkCk} 
 \end{split}
\end{align}
in terms of which the expectation values for the GFT energy-momentum tensor \eqref{eq:TABoperatorsNormal_saddPt} read 
\begin{align}
\begin{split}
     \expv{\T^{00}_k} = &   \frac{\sgn{\mkt}}{2 |\omega_{k/2}|} \left(\frac{k^2}{2  } \nk - m^2\ck \right)\,, \qquad 
   \expv{ \T^{a\neq b}_k} =  - \frac{\sgn{\mkt}}{8 |\omega_{k/2}|}k_a k_b \left( 2 \nk + \ck \right)\,, \\
   \expv{\T^{aa}_k} = &   \frac{\sgn{\mkt}}{2 |\omega_{k/2}|}\Bigg[ 
   -\left(2m^2 + \frac{k_a^2}{2}\right) \nk +  \left( \frac{k^2 - k_a^2 }{4}\right)\ck \Bigg]\,,\\
   \expv{ \T^{0b}_k} = &  \frac{\ii\, \sgn{\mkt} }{4  |\omega_{k/2}|} k_b \mathfrak{n}_k'\chiz \, .
    \label{eq:TsaddExp}
\end{split}
\end{align}
It then follows that $\frac{1}{3}{\rm tr} \expv{\T^{aa}_k} =  \frac{\sgn{\mkt}}{2 |\omega_{k/2}|}\Big[ 
   -\left(2m^2 + \frac{k^2}{6}\right) \nk +  \frac{k^2}{6}\ck \Big]$, which will be a useful expression in the following analysis.
For our choice of state, $\nk$ can be related to the expectation value of the number operator $N_k =   A_{k}^\dagger A_k$ (not to be confused with the lapse function), i.e., $\frac{\nk}{c_\sigma^2} = \expv{N_{k/2}}$. 
This relation is valid as long as $\sigma(\Vec{k})$ \eqref{eq:sigmaChoice} is symmetric in $k$ and we are within the range of validity of the saddle-point approximation. In particular, the exact form of $\sigma(\Vec{k})$ is irrelevant, as long as it is sufficiently peaked on the $\vec{k}=0$ mode.

To analyse the dynamics of the energy-momentum tensor components, 
 we first note that $\nk$ satisfies the equation of motion
\begin{align}
   \nk'' = 4 \omega_{k/2}^2 \nk\,.
\end{align}
As $\nk$ fully governs the dynamics of the squeezed energy-momentum tensor, the $\expv{\T^{AB}_k}$  satisfy similar dynamics, namely 
\begin{align}
\begin{split}
    \expv{\T^{00}_k} '' = & 4 \omega_{k/2}^2 \expv{\T^{00}_k} + \sgn{\mkt} 2 |\omega_{k/2}| m^2 \ck \,,\\
    \expv{\T^{a\neq b}_k} '' = &  4 \omega_{k/2}^2 \expv{\T_k^{a\neq b}} + \sgn{\mkt} \frac{k_a k_b}{2} |\omega_{k/2}| \ck\,, \qquad \\
    \expv{\T_k^{aa}} '' = &   4 \omega_{k/2}^2 \expv{\T_k^{aa}} - \sgn{\mkt} |\omega_{k/2}| \left( \frac{k^2 - k_a^2 }{2}\right)\ck\,, \\
      \expv{ \T_k^{0b}} '' = &  4 \omega_{k/2}^2 \expv{\T_k^{0b}}\,,
\end{split}
    \label{eq:TdynamicsSqueezed}
\end{align}
from which it follows that $\frac{1}{3}{\rm tr}\expv{\T_k^{aa}}'' = \frac{4}{3} \omega_{k/2}^2 {\rm tr}\expv{\T_k^{aa}} - \sgn{\mkt} |\omega_{k/2}| \frac{k^2}{3}\ck $. 
They also satisfy ($k_a \neq 0$)
\begin{align}
    \expv{\T^{00}_k}' = - \ii\, \frac{k^2}{k_a} \expv{\T_k^{0a}} , \qquad \expv{\T_k^{a\neq b}}' =  \ii\, k_a \expv{\T_k^{0b}}, 
    \qquad \expv{\T_k^{aa}}' =  \ii \,\left( 4 m^2 + k_a^2\right) \frac{\expv{\T_k^{0b}}}{k_b} \,  ,
\end{align}
where the index $b$ on the right-hand side of the last expression can refer to any space component of the energy-momentum tensor. 
Note in particular that due to the exponential growth of $\nk$, the constant terms in the expressions can be neglected at late times, leading to closed second-order equations for the $\expv{\T^{AB}_k}$ that are exactly those of the number operator.
 
The comparison of \eqref{eq:classicalCurr} and \eqref{eq:TsaddExp} allows an identification  regarding the nature of the perturbations focusing on the matching of factors of $k_a$; we see that the $\expv{\T^{AB}}$ resulting from our state choice are consistent with purely scalar perturbations. 
We first note that the overall factor of $k_b$ in $\expv{\T^{0b}}$ is consistent with vanishing vector modes $B^V_a = 0$. Similarly, from $\expv{\T^{a\neq b}}$ we find  $E_{a\neq b}^T = 0$ and $\pd_a E^V_b + \pd_b E^V_a = 0$ ($a\neq b$); we also get $\pd_a E^V_a = 0$ from $\expv{\T^{aa}}$. 
Finally, we conclude that $E_{aa}^T = 0$ by noticing that the $k_a^2$ terms in $\expv{\T^{aa}}$ give exactly the $k_a^2 E$ term in $j^{aa}$, using the identification $\sgn{\mkt}\expv{\T^{a\neq b}} = j^{a\neq b}$. 
The possibility of obtaining vector and tensor perturbations from the effective GFT metric we construct here is should be clarified in future studies; in what follows we  focus solely on scalar perturbations.

We can then use the above results and the relations found in \eqref{eq:pertQsols} to write down 
explicit expressions for the scalar metric perturbations arising from squeezed modes 
\begin{align}
\begin{split}
    E = &  
    \frac{1}{16 |\omega_{k/2}|}\frac{\pi_0 }{ a^4} (\ck+2 \nk)\,,\\
    B = & 
     - \frac{1}{4 |\omega_{k/2}| } \frac{1}{ a^2} \nk'\,,\\
    \psi =  & 
  \frac{1}{16 |\omega_{k/2}|} \left( 2\frac{m^2 }{\pi_0}\ck -  \left(k^2 \left( \frac{\pi_0}{a^4} + \frac{1}{\pi_0}\right) + \frac{4 m^2 \pi_0}{a^4} \right) \nk\right)\,,\\
    \tPhi = &  
   -\frac{1}{16 |\omega_{k/2}|}\left( \left( \frac{k^2 \pi_0}{a^4} - \frac{2 m^2 }{\pi_0}\right)\ck + \left( k^2 \left(-\frac{\pi_0}{a^4 } + \frac{1}{\pi_0}\right) - \frac{12 m^2 \pi_0}{a^4}   \right) \nk \right)\,.
     \label{eq:pertQsolsSqueezed}
\end{split}
\end{align}
Note that the overall sign factors in the explicit expressions for $\expv{\T^{AB}}$ cancel with the sign in the identification $j^{AB} = \sgn{\mkt} \expv{\T^{AB}}$, leading to simpler expressions.

From these effective expressions we can make some basic observations regarding the behaviour of  perturbations arising from squeezed modes:
\begin{itemize}
	\item 
	The initial spectrum of perturbations at the bounce, where we have $a^4 =  \pi_0^2$, can be computed as a function of $\vec{k}$. Since $\nk$ and $\ck$ as defined in \eqref{eq:defNkCk} scale as $e^{- \frac{\vec{k}^2}{4 s^2}}$, when $|\vec{k}|\gg s$ perturbations are exponentially small at the bounce. The parameter $s$ regulates the peakedness of the state \eqref{eq:sigmaGauss} and can be made arbitrarily small.
	On the other hand, we see that modes for which $|\vec{k}|\le s$ are of the same order as the background $\vec{k}=0$. 
	This differs from standard cosmological perturbation theory, where all perturbations are assumed to be small with respect to the background, and is a finite-width effect of the state we are considering: the situation of standard cosmology corresponds to the case of $s\to 0$ to obtain the background mode, and the inclusion of a separate spectrum for perturbation modes.
    In practice, only modes above a minimal $|\vec{k}|$ are observable, and very long-wavelength modes outside of that window could be absorbed into a redefinition of the background. Since in our case the equations of motion are linear and different $\vec{k}$ modes are decoupled, this would not introduce any nonlinear averaging effects.
	\item As the universe expands, $\nk$ increases and hence the perturbations grow in time. 
Even the combination $\nk/a^4$ grows (recall that at late enough times $a^4 \propto e^{2 m \chi^0}$ and $\nk \propto e^{2 |\omega_{k/2}| \chi^0}$), such that all perturbations grow  and take their minimum value at the bounce. 
	This fact can be reconciled with linear perturbation theory by recalling that the free GFT and the saddle-point approximation are applicable for a finite time only and furthermore, the perturbations are initially exponentially suppressed in $k$, i.e., the faster growing modes start with smaller initial amplitudes.	
	At late times, the terms proportional to $\frac{k^2}{\pi_0}\nk$ will be dominant in the expressions for $\tPhi$ and $\psi$ (assuming that the saddle-point approximation is still applicable). However, an approximation of the form $\tPhi \approx \psi \approx  - \frac{1}{16 |\omega_{k/2}|} \frac{k^2}{\pi_0}\nk $ would be invalid, as it violates \eqref{eq:combinedHGC}, which is derived directly from the harmonic gauge conditions. The harmonic gauge conditions are equivalent to the conservation law $\pd_0 \T^{0 B} + \ii \sum_a k_a \T^{aB} = 0$, which was  shown to hold exactly at operator level in \cite{gftmetric}. 
    \item In the $\vec{k}\to 0$ limit, $\psi $ and $\tPhi$ tend towards constants, as $\psi \sim \frac{1}{4 |\omega_{k/2}|} \left( \frac{m^2}{2 \pi_0}\ck - \frac{m^2 \pi_0}{a^4} \nk\right)$ and $\tPhi \sim \frac{1}{4 |\omega_{k/2}|} \left( \frac{m^2}{2 \pi_0}\ck + \frac{3 m^2 \pi_0}{a^4} \nk\right)$ and from the effective scale factor \eqref{eq:jTident} and the definition of $\nk$ \eqref{eq:defNkCk}, we can see that $|\omega_{k/2}| \to |m|$ and $\nk / a^4 \approx \text{const.}$
    These perturbations also satisfy the super-horizon limit of the harmonic gauge condition \eqref{eq:combinedHGC},  $\tPhi'' + 3 \psi'' = 0$.
	In the strict $\vec{k}\to 0$ limit, $E$ and $B$ do not appear in the metric where they are always multiplied by the wavenumber (or, equivalently, only enter as spatial gradients, see  \eqref{eq:classicalCurr}). 
	\end{itemize}

We proceed to analyse the concrete form of equations of motion for the perturbation variable $E$ arising for squeezed GFT modes and compare them to their classical counterparts.
Using \eqref{eq:generalEdynamicsTAB} and \eqref{eq:TdynamicsSqueezed}, the dynamics of $E$ can be written as 
\begin{align}
    E'' + 8 H E' + 4 ( H' + 4 H^2 - \omega_{k/2}^2) E + \frac{|\omega_{k/2}|}{4}\frac{\pi_0}{a^4}\ck = 0 \, .
    \label{eq:EdynSqueezed}
\end{align}
In the late-time limit we can neglect the $\ck$ term as it falls off as $a^{-4}$, and approximate $H' \approx 0$ and $H^2 \approx \frac{m^2}{4}$ (see \refsec \ref{sec:GFTuniverse_background}). If we also insert $\omega_{k/2}^2 = \frac{k^2}{4} + m^2$, we find
\begin{align}
    E'' + 8 H E'  - k^2 E  \approx 0\, .
\end{align}
This can be simplified further by considering an explicit late-time expression for $E'$. At late times, we can assume that $E \approx  \frac{\pi_0}{8 |\omega_{k/2}| a^4} \nk$, again neglecting the $\ck$ term, and $\nk' \approx  2 |\omega_{k/2}|\nk$ (see \eqref{eq:defNkCk}), leading to
\begin{align} 
    E' \sim -4 H E + 2 |\omega_{k/2}| E\, .
\end{align}
For small wavenumbers $\frac{k^2}{4}\ll m^2$ we furthermore have $|\omega_{k/2}| \sim 2 H$, such that $E' \sim 0$ and the equation of motion for $E$ simplifies to
\begin{align}
    E'' - k^2 E = 0\,.
    \label{eq:EeffectiveLate}
\end{align}
Comparing to \eqref{eq:pert_eomSingleFieldpert}, the corresponding equation in general relativity coupled to a single massless scalar field, our effective equation \eqref{eq:EeffectiveLate} has a Euclidean signature instead of the Lorentzian one, and is missing a factor of $\frac{\pi_0^2}{a^4}$.  
It furthermore resembles the general relativistic single field case \eqref{eq:pert_eomSingleFieldpert} more closely than that of four massless scalar fields \eqref{eq:Eeq}, which is similar to the results for the effective Friedmann equation as discussed in \refsec \ref{sec:GFTuniverse_background}.

In previous work in GFT cosmology (using different methods and assumptions) the signature of perturbations was found to depend on initial conditions, where both the Lorentzian and Euclidean case could be recovered \cite{Marchetti:2021gcv}. 
It is evident that alterations to the setup we present here are necessary to recover agreement with Lorentzian general relativity. The presence of a Euclidean signature for effective metric perturbations may not appear particularly surprising, given that our original GFT action (\ref{eq:GFTaction}) treats all four matter fields on the same footing and hence does not distinguish between ``space'' and ``time'' directions. A possibility which we will discuss in \refapp \ref{app:GFTaction} would hence be to start from a different coupling of matter degrees of freedom in the original GFT action.

In the general relativistic perturbation equations, the factor $\pi_0^2/a^4$ more generally reads $a^2/ N^2$ where $N$ is the lapse, and would hence be absent in the case of conformal time  $N \sim a$. 
The lapse is however determined by our choice of coordinate system and the expression of the conjugate momentum of the clock field, such that one would have to consider alternative matter actions to obtain a different form of $N$. 
As a particular example, one might want to consider k-essence models that include a more general function of the kinetic term in the Lagrangian for the four massless scalar fields. The challenge is then to obtain a model in which $N\sim a$ and $H^2 \sim $ const. at late times; we discuss an extension of our setup to k-essence models in \refapp \ref{sec:kEssence}. 
We note that the issue of a missing dynamical factor of $a^4/\pi_0^2$ was found in previous results on GFT perturbations \cite{Marchetti:2021gcv} and could be resolved through a more advanced construction 
\cite{Jercher:2023nxa, Jercher_2024_lett}, as we will discuss again in the conclusions. 

A discussion similar to the one we included for $E$ above could be carried out for the other three scalar perturbation variables. As these will generally suffer from similar deviations, we leave this analysis for future work. This concludes the analysis of squeezed modes. 

\subsection{Oscillating modes}
\label{sec:oscPert}

We now follow the equivalent procedure for oscillating modes $\omega_{k/2}^2 <0$ (see \refsec \ref{sec:GFT}).
From the definition $\omega_k^2 = m^2 + \vec{k}^2$ this will only hold for $m^2<0$ and for sufficiently small wavenumbers: in the saddle-point approximation we only consider the frequency $\omega_{k/2}$, which only corresponds to an oscillating mode for $k^2 < 4 |m^2|$.  
For the operator expectation values in \eqref{eq:TABoperatorsNormal_saddPt} and \eqref{eq:TABoperatorsNormal_saddPt_osc}, we obtain 
\begin{align}
\begin{split}
 \expv{ :A^{\dagger }_{k/2 }A_{k/2 }:} & =  \, \frac{1}{c_\sigma^2} e^{-\frac{k^2}{4 s^2}}(\ma^2 + \mb^2) =: \frac{1}{c_\sigma^2}\dk \, , \\
 \expv{:A^{\dagger }_{k/2 }\!^2:} + \expv{:A_{k/2 }\!^2:} & 
  = \, \frac{2 e^{-\frac{k^2}{4 s^2}}}{c_\sigma^2} 
\left( (\ma^2-\mb^2)\cos(2 |\omega_{k/2}|\chi^0) - 2\, \sgn{\mkt} \ma \mb \sin(2  |\omega_{k/2}| \chi^0) \right)\\
 & =: \frac{1}{c_\sigma^2}\mko \, ,
\label{eq:defDkMko}
\end{split}
\end{align}
which leads to  
\begin{align}
\begin{split}
    \expv{\T^{00}_k} \; = & \, \frac{\text{sgn}(\mkt)}{4 |\omega_{k/2}|}  \Bigg[ 4 |m^2| \dk
    + \frac{k^2}{2} \mko
   \Bigg]\,,\\
   \expv{\T^{0b}_k} \; = & \, -\frac{\ii k_b}{2} e^{-\frac{k^2}{4 s^2}} 
   \left[    \sgn{\mkt}(\ma^2 - \mb^2)\sin(2 |\omega_{k/2}|\chi^0) + 2 \ma \mb \cos(2 |\omega_{k/2}| \chi^0)
   \right] = \,\frac{\ii \, \sgn{\mkt} \,  k_b }{8 |\omega_{k/2}|} \mko'\,, \\
   \expv{\T^{a\neq b}_k} \; = &  \,  -\frac{\text{sgn}(\mkt)}{ 8 |\omega_{k/2}|} k_a k_b
   \left[2 \dk + \mko \right]\,,\\
    \expv{\T^{aa}_k} \; = & \,  \frac{\text{sgn}(\mkt)}{4 |\omega_{k/2}|} \Bigg[ (k^2 - k_a^2 )\dk 
    + \left(2|m|^2 -  \frac{k_a^2}{2}\right) \mko \Bigg]\,,
\label{eq:TABoperatorsNormal_osc_saddPt}
\end{split}
\end{align}
with $\text{tr} \expv{\T^{aa}_k}  = \frac{\sgn{\mkt}}{2 |\omega_{k/2}|} \left(  k^2 \dk + \left(3 |m|^2 - \frac{k^2}{4}\right)\mko \right) $.
The dynamics of oscillating modes are governed by $\mko$, which satisfies
\begin{align}
\mko'' = -4 |\omega_{k/2}|^2 \mko\,.
\end{align}
This leads to the following equations of motion for the GFT energy-momentum tensor: 
\begin{align}
\begin{split}
    \expv{\T^{00}_k} '' = & - 4 |\omega_{k/2}^2| \expv{\T^{00}_k} + \sgn{\mkt} 4 |\omega_{k/2}|  |m^2| \dk ,\\ 
    \expv{\T^{a\neq b}_k} '' = & - 4 |\omega_{k/2}^2| \expv{\T^{a\neq b}_k} - \sgn{\mkt} |\omega_{k/2}| k_a k_b \dk , \qquad \\
    \expv{\T^{aa}_k} '' = &   - 4 |\omega_{k/2}^2| \expv{\T^{aa}_k} + \sgn{\mkt} |\omega_{k/2}| \left( k^2 - k_a^2\right)\dk\,, \qquad  \expv{ \T^{0b}_k} '' =  - 4 |\omega_{k/2}^2| \expv{\T^{0b}_k}\,.
\end{split}
    \label{eq:Tdynamicsosc}
\end{align}
Note that this mimics the dynamical equations of squeezed modes \eqref{eq:TdynamicsSqueezed}, with an opposite sign, which hints at the possibility to recover a Lorentzian signature in the perturbation equations.

From \eqref{eq:pertQsols} and \eqref{eq:TABoperatorsNormal_osc_saddPt} we find the following expressions for perturbation variables in the case of oscillating modes:
\begin{align}
\begin{split}
    E = &  
    \frac{\pi_0 }{16 |\omega_{k/2}| a^4} (\mko+2 \dk)\,,\\
    B = & 
     -\frac{1}{8 |\omega_{k/2} | a^2} \mko'\,,\\
    \psi =& - \frac{|m^2|}{4 \pi_0 |\omega_{k/2}|} \dk + \frac{1}{8 |\omega_{k/2}|} \left( \frac{\pi_0 |m^2|}{a^4} - \frac{k^2}{4} \left( \frac{1}{\pi_0} + \frac{\pi_0}{a^4}\right) \right) \mko\,,\\
    \tPhi = & - \frac{1}{8 |\omega_{k/2}|} \left[  \left(\frac{2|m^2|}{\pi_0} + \frac{k^2 \pi_0}{a^4} \right) \dk + \left( \frac{k^2}{4} \left( \frac{1}{\pi_0} - \frac{\pi_0}{a^4}\right) + \frac{3  |m^2| \pi_0}{a^4} \right) \mko \right]\,.
     \label{eq:pertQsolsOsc}
\end{split}
\end{align}
Importantly, there are no growing terms in the perturbations, so that terms proportional to $\dk$ cannot be neglected at late times. 
The only applicable late-time limit is that the amplitude of terms proportional to $a^{-4}$ decreases. 
In particular, this implies that $E$ and $B$ decay, whereas $\psi$ and $\tPhi$ oscillate around a set value. Similarly to the case of squeezed modes, a late-time approximation in which $\tPhi \sim \psi$ violates the harmonic gauge condition \eqref{eq:combinedHGC} outside the super-horizon limit. In the $\vec{k} \to 0$ limit, the harmonic gauge condition reduces to $- \tPhi'' - 3 \psi'' = 0$ and is satisfied by the approximations $\tPhi \sim - \frac{|m^2|}{4 \pi_0 |\omega_{k/2}|}\dk - \frac{3|m^2| \pi_0}{8 |\omega_{k/2}| a^4} \mko$ and $\psi \sim - \frac{|m^2|}{4 \pi_0 |\omega_{k/2}|}\dk + \frac{|m^2| \pi_0}{8 |\omega_{k/2}| a^4} \mko$. In the late-time, superhorizon limit, we then find that $\psi \sim \tPhi$ are constants.

Using \eqref{eq:generalEdynamicsTAB} together with \eqref{eq:Tdynamicsosc} we find the following equation of motion for $E$ in the case of oscillating modes
\begin{align}
	E''  + 8 H E' + 4(4 H^2 +  H' + 4 |\omega_{k/2}^2| )E = \frac{|\omega_{k/2}| \pi_0}{2 a^4}  \dk\,.
\end{align}
The late-time limit is different from the squeezed case; in particular, the right-hand side is of the same order as $E$ and cannot be ignored. Hence the late-time limit for background quantities is less straightforward in the case of oscillating modes. As we discussed in \refsec \ref{sec:GFTuniverse_background}, oscillating modes do not lead to a bouncing universe at the background level, so that we would assume that oscillating modes appear only in conjunction with at least one squeezed mode that gives desirable background dynamics. In such a case, we would have $ 4 H^2 \sim m_{\rm sq}^2$ and $H' \sim 0$, where $m_{\rm sq}^2$ would be the value of $m^2$ for the squeezed background mode, which is different from the value appearing in $\omega_{k/2}^2$.
While we recover a Lorentzian signature, the discrepancy of the $a^4/\pi_0^2$ factor remains. 
 Lastly, as no terms can be neglected at late times, $E'$ cannot be simplified and we are left with additional terms compared to general relativity \eqref{eq:Eeq}. As an additional point of comparison, we noted below (\ref{eq:Eclassical}) that in the case of general relativity with a single massless scalar field, the amplitude of $E$ falls off as $1/a$ with increasingly rapid oscillations. This is clearly very different from the $1/a^4$ fall-off with a constant oscillation frequency that we observe in the explicit solution \eqref{eq:pertQsolsOsc}.

This concludes the analysis of scalar perturbations within our proposal to extract an effective metric from GFT for a first naive state choice. 
We have obtained explicit expressions for scalar perturbations in the squeezed and oscillating case. In both cases, we find that the equation of motion for the effective perturbation variable $E$ shows deviations from the general relativistic dynamics in the form of having the wrong signature (squeezed modes), having additional terms (oscillating modes), and a missing dynamical factor of $a^4/\pi_0^2$ (both cases). 
In a model with at least two modes, one oscillating and one squeezed, one could imagine that the squeezed mode is very highly peaked on the background mode and leads to a bouncing universe, while the perturbative modes are suppressed, and the oscillating mode gives the dominant contribution to cosmological perturbations. Still, in order to match the general relativistic dynamics at late times, alterations to the proposed setup, e.g., in form of a more complicated state choice, are required. Below we will discuss some possible directions for obtaining more phenomenologically acceptable results in our general setting.

\section{Conclusion}\label{sec:conclusion}

In this paper we extended the analysis of an effective metric for an FLRW background as studied in \cite{gftmetric} to cosmological perturbations.
For the FLRW background, previous work had shown the promising result  of an exactly flat metric and a bouncing universe. The effective Friedmann equation derived for this case showed agreement with general relativity coupled to a single massless scalar field, while the GFT model includes four such fields.
The study of GFT perturbations then has two main objectives: firstly, to establish whether the effective metric proposal allows a consistent reconstruction of perturbation variables, and secondly to investigate whether their effective dynamics can be interpreted from the perspective of general relativity (in suitable limits).
 
We began with a brief summary of the basics of GFT and the fundamental ideas behind the effective metric proposal and its application to homogeneous cosmology.
As in most of the literature on GFT cosmology, we neglect interactions between GFT quanta. The free theory then contains two types of modes, oscillating and squeezed modes, which exhibit different dynamics.
Throughout the paper we work in a relational coordinate system given by four massless scalar fields. The proposal is then to identify the GFT energy-momentum tensor with an effective spacetime metric, given that both define conserved Noether currents associated to the same symmetries. 
By giving explicit expressions for the general relativistic Noether currents for a perturbed FLRW metric, we were able to reconstruct expressions for all scalar metric perturbations explicitly for the first time in the GFT literature. This is a significant improvement on previous work in which only the spatial volume and its perturbations could be studied.
We established general expressions and equations of motion for scalar perturbation variables in terms of the effective operator dynamics, which are independent of a specific state choice. 
These could be used for any GFT state beyond the example we consider here, or for more general models with alternative operator dynamics, e.g., for a different GFT action.
While the proposal for an effective GFT metric  is very  general, in the sense that an effective metric can be associated to any state that is sufficiently semiclassical, the particular choice of state governs the specific form of such a metric and its symmetries.

For our analysis, we chose a Fock coherent state highly peaked on the homogeneous  $\vec{k}=0$ mode. 
Fock coherent states are commonly used in the GFT literature as they satisfy the requirement of semiclassicality; peaking around the homogeneous mode reflects the goal of obtaining an FLRW metric with small perturbations.
This state was used in \cite{gftmetric} to obtain the effective background metric, and its non-zero $\vec{k}$ modes are interpreted as perturbations.
In \cite{gftmetric} we had shown that the effective Friedmann equation for squeezed modes corresponds (at late times) to what is expected for general relativity coupled to a single massless scalar field.
GFT is not a direct quantisation of classical general relativity; a GFT action is constructed via symmetries, renormalisation arguments and connection to discrete quantum gravity models. Our introduction of a specific simple coherent state and the truncation to the free theory are also significant simplifications. Hence, while obtaining a reasonable Friedmann equation gives a first hint, it does not yet give strong evidence that the resulting low-energy theory is consistent with general relativity.

To obtain explicit expressions for perturbative quantities we made use of the saddle-point approximation, which restricts the validity of our results to perturbations with sufficiently small wavenumbers and to a finite time. 
For our state choice small wavelength perturbations are initially exponentially suppressed and the assumption of negligible interactions in the GFT action limits our results to regions close (enough) to the bounce. 
We considered the case of oscillating and squeezed modes separately, where for both mode types the dynamics of perturbations are naturally very similar to those for the respective background mode.
Our choice of state leads to expressions for the expectation values of the GFT energy-momentum tensor components $\expv{\T^{AB}}$ that are compatible with the interpretation of recovering only scalar perturbations, even though in principle the components of $\expv{\T^{AB}}$ contain all perturbation types. 

The effective perturbations we found for squeezed modes grow in time, excluding a consistent interpretation as small deviations from a homogeneous background at a certain point in the evolution. 
Comparing the equation of motion for the perturbation variable $E$ to those obtained in general relativity for either one or four massless scalar fields revealed several discrepancies. Firstly, the dynamics of the effective perturbation have a Euclidean signature instead of a Lorentzian one. Secondly, they resemble the general relativistic dynamics one might expect for conformal time, whereas we are working in a harmonic gauge given by the relational coordinate system, which would lead to a relative factor $a^4/\pi_0^2$. Finally, the late-time limit of the effective dynamics for $E$ resemble (save for the aforementioned discrepancies) those of general relativity with a single matter field, which is similar to what we found for the effective Friedmann equation. 
To recover a bouncing universe at the background level, we saw that at least one squeezed mode needs to be excited;  for our state choice we then inevitably encounter perturbations that are of squeezing type. 
Oscillating modes, on the other hand, remain finite in amplitude throughout the evolution of the universe. While for these modes we recovered a Lorentzian signature in the dynamical equations for the effective perturbation $E$, we again encountered the same discrepancy regarding a dynamical factor of $a^4/\pi_0^2$, moreover, additional terms that are not present in general relativity arise. Note that since a single oscillating mode does not lead to an expanding universe at the background level, one needs to consider a minimum of two $J$ modes (one squeezed, one oscillating) in order to have perturbations of oscillating type in a phenomenologically feasible universe.

The GFT literature includes models for Euclidean as well as for Lorentzian gravity. As we discussed in our earlier review of GFT, the desired spacetime signature can (but does not need to) be built into a choice of gauge group; to determine the relation between such details of the model and the emergent spacetime signature, one needs to have access to an effective spacetime geometry.
The work of \cite{gftmetric}, as well as other results in GFT cosmology \cite{Marchetti:2021gcv}, indicated that in GFT the metric signature is not fundamentally included in the quantum theory and instead emerges at an effective level. 
In our case, at the background level the metric signature (read off from effective metric coefficients) is determined by initial conditions, whereas for the perturbations it is determined by the type of dynamical equation and depends on the mode type. 
Since we are working in a free GFT, the model presented here has no coupling between the background and perturbations, and the dynamics of the various $\vec{k}$ modes of the GFT energy-momentum tensor are independent of one another.  This is a reflection of the linearity of the theory and stands in contrast to general relativity, where the perturbation equations explicitly depend on the scale factor and the Hubble rate. This might suggest that the proposed setting is more suitable for studying perturbations around a flat spacetime, the study of which we leave to future investigations.

All previous approaches to cosmological perturbations in GFT were limited to considering the perturbation of the volume element and thus the combination $k^2E - 3 \psi$. The exceptional advantage of having access to a reconstructed metric in our setting lies in the fact that we can retrieve any combination of perturbative quantities, in particular, we can construct effective gauge-invariant perturbations. 
This is of particular interest as gauge-invariant quantities are those that can be related to observations. Moreover, future work based on more elaborate state choices could include tensor or vector modes, which could again be read off from the effective metric.

In a previous study \cite{Marchetti:2021gcv} based on volume perturbations, the dynamics of perturbations similarly lack a factor of $a^4/\pi_0^2$, whereas the signature is determined by initial conditions. The model is also built on a free GFT action with a single group field, but includes a fifth matter field that is assumed to dominate the relational fields. One hence assumes agreement with general relativity coupled to a single scalar field, which is found at late times in the $\vec{k} \to 0 $ limit, similar to the long-wavelength limit of the dynamics for $E$ we find here in the case of squeezed modes.
The dynamical discrepancies in the perturbation equations at finite $\vec{k}$ could then be resolved in \cite{Jercher:2023nxa} using a GFT model with two types of group field (``spacelike'' and ``timelike''), which enter the GFT action in a different manner. In this work the choice of state is not a simple coherent state, but includes entanglement between the perturbations of the different GFT fields. This choice of state and a simplified form of GFT dynamics assumed in the analysis allow for the introduction of a free function that can be chosen to achieve agreement with general relativity at late times, including fixing the signature to Lorentzian.
The mismatch we find between effective GFT dynamics and those of general relativity is similar to those of  \cite{Marchetti:2021gcv}, so the follow-up work of \cite{Jercher:2023nxa} might suggest that this can be improved in a more complicated GFT and with a different choice of state.

In the appendix we propose some avenues for extending our results or their interpretation in terms of a corresponding classical theory.
Beyond this, our results for cosmological perturbations can be extended in many different directions:
\begin{itemize}
\item Alternative state choices. 
The state we considered here is characterised by a single $\vec{k}$-dependent function that determines both background and perturbations. One might want to consider states that more closely resemble the approach in standard cosmology, e.g., one could consider a background mean field only used to define background quantities, plus a small $\vec{k}$-dependent contribution for the perturbations, which can exhibit an entirely different spectrum. 
 This would be similar to previous  studies of perturbations in GFT \cite{Gerhardt:2018byq, Marchetti:2021gcv, Jercher:2023nxa}.
\item Improved GFT dynamics. 
The dynamics of perturbations are derived from dynamics of the GFT energy-momentum tensor, which are determined by the GFT action.
Including the effect of interactions in the GFT action, which we have neglected here, might lead to effective dynamics that are closer to those of general relativity. This would be consistent with the origin of GFT in discrete gravity models, where the precise form of interaction is important to determine the ``gluing'' of lower-dimensional building blocks to form spacetime. Viewed from that angle it would seem unreasonable to expect that a truncation to a simple quadratic action that does not know about interactions is already able to capture general-relativistic dynamics (such a result would also suggest that almost any GFT action of a particular class reduces to general relativity, which might again seem unrealistic). Of course, adding interactions will substantially complicate matters and potentially require new (perturbative and nonperturbative) methods.
\item Including additional group fields. 
One might consider extensions similar to \cite{Jercher:2023nxa}, where two GFT fields are included, whose interplay leads to dynamical equations that agree with GR in a certain limit.
Such extensions would change the form of the $\T^{AB}$ and symmetry requirements of the energy-momentum tensor would likely impose certain conditions (and possibly limitations) on such a construction. 
\end{itemize}

Finally, we emphasise that the setup for reconstructing an effective GFT metric introduced in \cite{gftmetric} is general and not limited to  cosmology. Its usefulness could be established by investigating its application also outside of the context of homogeneous and isotropic cosmology. 
Here, anisotropic Bianchi models might be best suited and black hole spacetimes would be of particular phenomenological interest. 
If proven suitable for obtaining a variety of spacetimes, the effective GFT metric could pave the way for a variety of fruitful future research directions.

\acknowledgments
The work of SG is funded by the Royal Society through the
University Research Fellowship Renewal URF$\backslash$R$\backslash$221005. The work of LM was partly funded by the Leverhulme Trust through a Study Abroad Studentship.

\appendix

\section{Possible extensions}
\label{app:GFTuniverse_comments}

In our analysis of the effective GFT metric recovered for the cosmological setting, the comparison with general relativity gave two main results: 
\begin{enumerate}
\item The effective Friedmann equation obtained at late times disagrees with that of general relativity with four massless scalar fields. Instead, consistent with previous literature, the Hubble rate (defined for the clock $\chi^0$) approaches a constant, which resembles the case of a single scalar field. This discrepancy was discussed in detail in \cite{gftmetric}.
\item Effective dynamics of perturbations do not agree with those of Lorentzian general relativity; we find a Euclidean signature for effective perturbations for squeezed modes, and a factor of $a^4/\pi_0^2$ is missing from the equations of motion.
\end{enumerate}
In order to recover a suitable semiclassical regime for cosmology, alterations have to be introduced to the setup described above. 
In this appendix we consider two routes to such alterations that focus on the manner in which the scalar fields are included in the theory and demonstrate the restrictions imposed by the setup. We first consider changing the GFT action and coupling clock and spatial scalar fields differently. Then, we consider the possibility of comparing not with general relativity with four free massless scalar fields, but with the more general setting of k-essence models, keeping in mind that GFT does not arise from quantising a particular classical matter action but relies on general symmetry arguments.
We find that both cases are restricted by symmetry requirements on the form of the GFT energy-momentum tensor and the conserved classical currents. 
We present both considerations separately; the combination of both approaches (comparing more general GFT actions to k-essence models) is left for future work.

\subsection{Extensions of the GFT action}
\label{app:GFTaction}

In the construction of a GFT action for quantum gravity with four massless scalar fields, one imposes some symmetries of the corresponding classical action (\ref{eq:GFT_scalaractionFourFields}), namely, shifts, rotations and reflections. 
The Laplacian on $\mathbb{R}^4$ in the GFT action \eqref{eq:GFTaction} is consistent with these; in particular, derivatives with respect to the scalar fields all enter with the same prefactor to preserve the rotational symmetry under $\chi^A\rightarrow {O^A}_B\chi^B$. 
As the $E(4)$ symmetry is broken upon singling out a clock field for quantisation in the deparametrised approach to GFT, one might want to impose an $E(3)$ symmetry between the spatial fields only and allow for a different factor in front of the derivatives with respect to the clock field, as was considered already in \cite{Gielen:2017eco}. 
Introducing a new parameter $c_a \in \mathbb{R}$, this leads to a more general form of the free action
\begin{align}
    S = \int \text{d}^4 \chi\, \lagr , \qquad \lagr= \sum_J\left(\frac{1}{2}\mathcal{K}_J^{(0)}\varphi_J^2 - \frac{1}{2}\mathcal{K}_J^{(2)}\left( (\pd_0\varphi_J)^2 +c_a\sum_a(\partial_a\varphi_J)^2\right)\right)  \, .
    \label{eq:extendedGFTaction}
\end{align}
The action \eqref{eq:GFTaction} used in our analysis so far is evidently recovered for $ c_a  =1$. On the other hand, setting $c_a=-1$ means the Laplacian is now the one on Minkowski spacetime $\mathbb{R}^{3,1}$ and the symmetry group $E(4)$ is replaced by the Poincar\'e group $E(3,1)$, which one might hope could encode Lorentzian rather than Euclidean signature in the effective spacetime geometry.

Having introduced a Lorentzian structure on the space spanned by the four massless scalar fields, the energy-momentum tensor thus obtains a non-trivial index structure and we adjust its definition to
\begin{align}
    \tensor{T}{^A_B} := - \frac{\pd \lagr}{\pd(\pd_A \varphi)} \pd_B \varphi + \lagr\, \delta^A_B\,,
\end{align}
where indices are raised and lowered with the Minkowski metric $\eta_{A B} = \text{diag}(-1, \, 1, \, 1,\, 1)$. 
Specifically, $T^{AB}$ is symmetric (unlike $\tensor{T}{^A_B}$) and therefore suitable for an identification with $j^{AB}$ as defined in \eqref{eq:jmetric}. Note that for $c_a \neq 1$ and the previous definition of the GFT energy-momentum tensor \eqref{eq:TGFTdef}, $T^{AB}$ would no longer be symmetric and a consistent identification with the classical currents impossible.

The question is then how such a change affects phenomenology.
The additional factor in front of the spatial gradient term enters the definition of $\omega_k$ in the GFT Hamiltonian \eqref{eq:hamiltonian} via $\omega_k^2 = m^2 +\vec{k}^2 \ \to \ \omega_k^2 = m^2 + c_a \vec{k}^2$.
Recall that we obtain a squeezing-type Hamiltonian for $\omega_k^2 >0$, while $\omega_k^2 < 0$ results in a harmonic oscillator Hamiltonian.
If we consider the case with $c_a  = -1$, we find that all modes with $m^2 <0$ are oscillating modes. For $m^2 >0$, we find squeezed modes only in the cases where $\vec{k}^2 < m^2$; all other modes are of oscillating type. 
Such a theory might be phenomenologically desirable as it has a limited number of squeezed (exponentially growing) modes.  

To assess possible effects on the signature of the effective metric, we first note that due to the necessary raising of the index of $\tensor{T}{^0_0}$, we find an additional minus sign in the effective expression, whereas the spatial diagonal components remain unchanged. 
For a squeezed background mode, we thus find (see \eqref{eq:jTident}) $\expv{\T^{00}_0} = - \sgn{\mkt} |m| (\mb^2 - \ma^2) \propto j^{00} = \pi_0$ and $\expv{\T^{aa}_0} = - \sgn{\mkt} |m| \mathfrak{n}_0(\chi^0) \propto j^{aa} = -\frac{a^4}{\pi_0}$. 
Recall that at the level of the background the effective metric signature is determined by the signs of $\expv{\T^{00}_0}$ and $\expv{\T^{aa}_0}$ (see \eqref{eq:jAB_bg}): If all components of $j^{AB}$ have the same sign the metric is Euclidean, if the $j^{00}$ component has a different sign, the metric is Lorentzian. 
In order to recover a  Lorentzian signature for the background metric we therefore need $\sgn{\mb^2 - \ma^2} \neq \sgn{\mathfrak{n}_0(\chi^0)} = 1$, i.e., we require $\ma^2> \mb^2$ for the initial conditions. Using the same identification as before, $j^{AB} = \sgn{\mkt} \expv{\T^{AB}}$,  we then recover a positive $\pi_0$ from the effective expressions. In short, setting $c_a =-1$ in the GFT action has implications for the range of initial conditions that give a Lorentzian effective FLRW metric.

In the case of effective perturbations, the signature depends on their dynamics, which again are determined by the mode type. As pointed out above, which modes are of squeezed or oscillating type changes with $c_a=-1$. 
Additionally, the sign of $\expv{\T^{00}_k}$ has changed, which enters the explicit forms of $\psi$ and $\tPhi$, see \eqref{eq:pertQsols}. The overall dynamics of squeezed or oscillating perturbation modes should however remain unaffected and we should find the same discrepancies in the dynamics as noted in the main text.\\

Finally, we comment on more general extensions of the GFT action:
To uphold the premise of our proposal, any generalisation of the construction presented must lead to a symmetric GFT energy-momentum tensor that can be consistently identified with the classical currents. 
If one wanted to extend the GFT action to include, e.g., higher-order derivatives, such higher-order terms must then appear for all four scalar fields. Such a modification will inevitably also affect the background dynamics and one cannot include additional terms solely for the spatial fields, which might have been desirable from a purely phenomenological perspective.

\subsection{K-essence models}
\label{sec:kEssence}

In the main text, we compared the effective GFT dynamics to dynamics of free massless scalar fields in general relativity and found various disagreements. However, recall that GFT is not a direct quantisation of any classical theory but constructed from general principles and properties of Feynman amplitudes, to be interpreted as discrete spacetime histories. The GFT action we discussed uses the shift symmetries of free massless scalar fields, but such symmetries exist in more general classical matter theories. It would hence be very reasonable to suggest that the classical limit of GFT could correspond to such a more general matter theory. In the following, we investigate this proposal and study a more general form of the scalar field action known as k-essence models (see, e.g., \cite{kEssence1, kEssence2}).

When using four scalar fields as a relational coordinate system, as done here and in previous GFT works \cite{Gielen:2020fgi, Marchetti:2021gcv, Jercher:2023nxa, gftmetric}, the simplest assumption is to assume the Lagrangian for free, minimally coupled fields
\begin{align}
    \lagr_\chi = - \frac{1}{2}\sqrt{-g}\sum_A g^{\mu \nu}\pd_\mu \chi^A \pd_\nu \chi^A\,.
    \label{eq:lagrSimple}
\end{align} 
However, the construction of an effective GFT metric only requires  a shift symmetry under $\chi^A\rightarrow \chi^A+\epsilon^A$, which is satisfied by any Lagrangian that only depends on derivatives of the scalar fields. In particular, we can generalise to a Lagrangian of the form ($a = 1,\, 2,\, 3$)
\begin{align}
    \lagr = \sqrt{-g}\, \mmp(X_0, X_a), \quad \text{with}  \quad X_0  = -\frac{1}{2} g^{\mu \nu} \pd_\mu \chi^0 \pd_\nu \chi^0\,, \; X_a = -\frac{1}{2} g^{\mu \nu} \pd_\mu \chi^a \pd_\nu \chi^a\,,
    \label{eq:matterLagrStandard}
\end{align}
where $\mmp$ denotes a general function. 
For a flat FLRW spacetime and in a relational coordinate system with $\pd_\mu \chi^A = \delta_\mu^A$ we have  $X_0 = \frac{1}{2 N^2}$ and $X_a = - \frac{1}{2 a^2}$. 
For a Lagrangian as given in \eqref{eq:matterLagrStandard},
the energy-momentum tensor is given by\footnote{We do not carry out the sum explicitly in the last step to avoid a confusing index structure.}
\begin{align}
    ^{(\chi)}\!\tensor{T}{^\mu_\nu} = \delta^\mu_\nu \, \mmp + \sum_A \frac{\pd \mmp}{\pd X_A}g^{\mu\alpha} \pd_\alpha \chi^A \pd_\nu \chi^A  =  \delta^\mu_\nu \, \mmp + \sum_A \frac{\pd \mmp}{\pd X_A}g^{\mu A} \delta^A_\nu \,,
    \label{eq:matterT}
\end{align}
and the classically conserved currents in relational coordinates read
\begin{align}
    (j^\mu)^A =- \sqrt{-g} \,\frac{\pd \mmp}{\pd X_A}g ^{\mu A} \quad \text{(no sum over $A$)} \,.
    \label{eq:kessenceCurrents}
\end{align}
The clock field momentum is given by
\begin{align}
    \pi_0 = \frac{\sqrt{|q|}}{N}\frac{\pd \mmp}{\pd X_0}\,,
    \label{eq:lapseFixingKessence}
\end{align}
which upon fixing $\mmp$ gives an equation that can be solved for the lapse $N$. 

In the following we demonstrate with a simple example that we could obtain a constant Hubble rate within k-essence models; we will discuss the restrictions imposed by symmetry requirements below.
For the example we choose $\mmp = (X_0)^u + \sum_a (X_a)^v$, with $u,\, v \in \mathbb{R}$.
If we assume that, again for a flat FLRW universe, $N = \xi\, a^q$ for some $\xi \in \mathbb{R}^+$ and $q \in \mathbb{R}$, we obtain the following expression for the conserved canonical momentum of the clock field and a relation between $q$ and $u$:
\begin{align}
    \pi_0 \propto a^{3 + q(1- 2u)} \overset{!}{=} \text{const.} \qquad \Rightarrow 
\qquad q = \frac{3}{2u -1}\,.
\label{eq:qcondition}
\end{align}
Note that an explicit form of $\xi$ as a function of $\pi_0$ can be found from \eqref{eq:lapseFixingKessence}. 
From \eqref{eq:matterT} we can obtain the energy density 
\begin{align}
    \rho =  -\mmp - \frac{\pd \mmp}{\pd X_0} g^{00}  =  \frac{1}{ 2^uN^{2u}} \left(2u - 1\right) + (-1)^{v+1}   \frac{3}{2^v a^{2v}}\,,
\end{align}
leading to the Friedmann equation
\begin{align}
    H^2 \propto &  \, N^2 \rho =  \frac{\left(2u - 1\right)}{ 2^u \xi^{2(u-1)}} \frac{1}{a^{2q(u-1)}}  + (-1)^{v+1}   \frac{3 \xi^2}{2^v } \frac{1}{a^{2(v-q)}} \,,
\end{align}
where we again assumed that $N = \xi\, a^q$.
We would then recover a constant right-hand side of the Friedmann equation in the case of $u =1$ and $v = q$, with $q = 3$ from \eqref{eq:qcondition}, so that we recover a harmonic gauge. 
The discrepancy between the GFT effective Friedmann equation and the general-relativistic Friedmann equation at late times for four massless scalar fields would be resolved in this type of model.
In the case of conformal time ($q=1$), we have $u = 2$, such that the first term in the Friedmann equation decays as $a^{-2}$, and the right-hand side approaches a constant at late times for $v = 1$. Note however that this would imply a stronger mismatch with the GFT Friedmann equation \eqref{eq:effFried}, which does not contain an $a^{-2}$ term, at early times.

So far, we have focused solely on the homogeneous FLRW dynamics, where the off-diagonal parts of $(j^\mu)^A$ vanish. 
If we consider the time-space components of the currents \eqref{eq:kessenceCurrents}, we find
\begin{align}
    (j^0)^a = - \sqrt{-g} g^{0a}\frac{\pd \mmp}{\pd X_a}, \qquad  (j^a)^0 = - \sqrt{-g} g^{0a}\frac{\pd \mmp}{\pd X_0}\,,
\end{align}
and in order to relate these to a symmetric GFT energy-momentum tensor we must demand 
$(j^{0})^a = (j^a)^0$. 
This imposes $\frac{\pd \mmp}{\pd X^0} = \frac{\pd \mmp}{\pd X^a}$, and the Lagrangian has to include all fields in the same manner.
While this symmetry requirement only applies when considering non-diagonal metrics, one might prefer a general construction that can hold for various spacetimes (in particular, also for the perturbed FLRW case). With this restriction we have
\begin{align}
\begin{split}
    \lagr = & \sqrt{-g}\, \mmp(X) \ \ \text{with} \ \ X  := -\frac{1}{2} \sum_A g^{\mu \nu} \pd_\mu \chi^A \pd_\nu \chi^A = \frac{1}{2 N^2} - \frac{3}{2 a^2}\,, 
    \quad (j^\mu)^A = - \sqrt{-g} g ^{\mu A} \frac{\pd \mmp}{\pd X}  \, ,
    \label{eq:symmRequirement}
\end{split}
\end{align}
where we assumed a flat FLRW background and a relational coordinate system for the explicit form of $X$.
The requirement that 
\begin{align}
\pi_0 = \frac{a^3}{N}\frac{\pd \mmp}{\pd X} \overset{!}{=} \text{const.}
\end{align} 
leaves us with two possible scenarios:
\begin{enumerate}
    \item The case we considered in the main text above, i.e., $N \propto a^3$ and $\mmp \propto X$.
    \item (Almost) conformal time: $X$ is a single power of $a$, i.e., $N = \xi \, a$ , $\xi \in \mathbb{R}^+$, such that $X = \frac{1}{2a^2} \left(\frac{1}{\xi^2} - 3 \right)$. We then have $\pi_0 = \frac{a^2}{\xi} \frac{\pd \mmp}{\pd X} \overset{!}{=} \text{const.}$ and therefore $\mmp \propto X^2$. 
    We then find $\pi_0 \propto \frac{1}{\xi^3} - \frac{3}{\xi}$
    which can be solved to obtain an expression for $\xi (\pi_0)$. As $\pi_0$ is positive, we must have $\xi \in (0,1/\sqrt{3})$.
\end{enumerate}

From the energy density $- \rho  = \mmp + \frac{\pd \mmp}{\pd X} g^{00}$ we can again compute the Friedmann equation $H^2 \propto -N^2 \mmp + \frac{\pd \mmp}{\pd X} $.
The first option above is the standard case we considered in the main text and results in \eqref{eq:clFried}. The second case of conformal time gives ($\mmp \propto X^2$)
\begin{align}
    H^2 \propto  \frac{3}{4 a^2} \left( \frac{1}{\xi^2}  -3 \xi^2 -2  \right)\,,
\end{align}
which is positive as $\xi^2 < \frac{1}{3}$ (see above).
This goes to zero at late times and does not match the GFT effective Friedmann equation. 
Hence, with the symmetry restriction that enables a consistent identification of the classical currents with the GFT energy-momentum tensor the scope of allowed k-essence models is limited and we could not identify a case in which we recover a constant Friedmann equation at late times. 

\subsection{Summary}

To obtain effective GFT dynamics that can match those of general relativity at the late times, alterations to the setup presented in the main text are necessary. These can go in several directions: one can either change the definition of the GFT model, or the classical matter theory one expects to obtain at low energies. We studied these two types of modifications separately. Both cases are limited by symmetry requirements on $T^{AB}$ and $j^{AB}$, which are crucial to allow a consistent identification with one another. Hence, the desirable effects  we found cannot straightforwardly be included in the setup of an effective GFT metric as studied in the main text.

Adjusting the derivatives with respect to clock and spatial fields in the GFT action has the potential to alter which values of $\omega_k$ result in oscillating or squeezed modes. Interestingly, it is possible to introduce a maximum wavenumber for squeezed modes, such that all modes with larger $k$ will be of oscillating type. 
In the case of k-essence models,
 we assessed whether it is possible to find a form of the matter Lagrangian that gives conformal time and a constant general relativistic Friedmann equation at late times.
The desired result can be achieved if one is concerned solely with an FLRW metric; then we saw that clock and rod fields can be included in the classical k-essence action in such a way that one recovers a constant Hubble rate in general relativity with four massless scalar fields, thus matching the result of GFT. However, the goal of our paper was to explicitly include cosmological perturbations, and one would like to find a common consistent description for background and perturbations in which the phenomenology of both is satisfactory from the perspective of general relativity. 

\bibliographystyle{ieeetr}

\bibliography{bib_part2}

\end{document}